\documentclass[prc,amsmath,amssymb,superscriptaddress,showpacs,preprint,
tightenlines]{revtex4}

\usepackage{graphicx,dcolumn,epsfig}

\begin{document}

\title{Instantaneous Shape Sampling - a model for the $\gamma$-absorption
cross section of transitional nuclei}

\author{I.\,Bentley}
\affiliation{Department of Physics, University of Notre Dame,
             Notre Dame, IN 46556, USA}
\author{S.\,Brant}
\affiliation{Department of Physics, Faculty of Science, University of Zagreb, 
             10000 Zagreb, Croatia}
\author{F.\,D\"onau}
\affiliation{Institut f\"ur Strahlenphysik,
             Forschungszentrum Dresden-Rossendorf, 01314 Dresden, Germany}
\author{S.\,Frauendorf}
\affiliation{Department of Physics, University of Notre Dame,
             Notre Dame, IN 46556, USA}
\author{B.\,K\"ampfer}
\affiliation{Institut f\"ur Strahlenphysik,
              Forschungszentrum Dresden-Rossendorf, 01314 Dresden, Germany}
\author{R.\,Schwengner}
\affiliation{Institut f\"ur Strahlenphysik,
             Forschungszentrum Dresden-Rossendorf, 01314 Dresden, Germany}
\author{S.\,Q.\,Zhang }
\affiliation{School of Physics and State-Key Laboratory of Nuclear Physics and 
             Technology, Beijing University, Beijing 100871, PR China}
\date{\today}

\begin{abstract} The influence of the quadrupole shape fluctuations on the
dipole vibrations in transitional nuclei is investigated in the framework of
the Instantaneous Shape Sampling  Model, which combines the Interacting
Boson Model for the slow collective quadrupole motion with the Random
Phase Approximation for the rapid dipole vibrations. Coupling to the
complex background configurations is taken into account by
folding the results with a Lorentzian with an energy dependent width. 
The low-energy energy portion of the  $\gamma$- absorption 
cross section, which is important for photo-nuclear processes, is studied for
the isotopic series of  Kr, Xe, Ba, and Sm. The experimental
cross sections are well reproduced. The low-energy
cross section is determined by the Landau fragmentation of the dipole strength
and its redistribution caused by the shape fluctuations.
Collisional damping only wipes out fluctuations of the absorption 
cross section, generating the smooth energy dependence observed in experiment.  
In the case of semi-magic  nuclei, shallow pygmy resonances  are found in agreement with experiment.
 \end{abstract}


\maketitle

\section{ Introduction}
\label{intro}

The cross sections of photo-nuclear processes, such as ($\gamma,n$),
($n,\gamma$), ($\gamma,p$), ($p,\gamma$), ($\gamma,\alpha$), ($\alpha,\gamma$)
are key elements in various astrophysical scenarios, like  supernovae
explosions or $\gamma$-ray bursts. Precise values of these basic data are also
indispensible for simulations of processes of nuclear technology. Many of the
relevant reactions involve unstable nuclei for which measurements of the cross
sections are not possible. Therefore, the cross sections have to be taken from
theory. In many cases the statistical Hauser-Feshbach model is applicable,
which decomposes the total reaction cross section into a product of the
absorption and emission probabilities of the particles and $\gamma$-quants.
Theoretical models that predict  the dipole strength
function for $\gamma$-absorption or emission through the whole nuclear chart
are therefore of utter importance. The reactions take place in an energy
interval of a few MeV around the particle emission thresholds. Aside from the mentioned
applications in nuclear astrophysics and nuclear technology, the understanding
of the mechanisms that determine the structure of the dipole strength function 
on the low-energy tail of the isovector giant dipole resonance is a
challenge of its own to nuclear theory. 

The present article proposes and tests a new approach, which we call
Instantaneous Shape Sampling (ISS). This approach combines the microscopic
Quasiparticle Random Phase Approximation  for dipole excitations with the
phenomenological Interacting Boson Approximation  for a dynamical
treatment of the nuclear shape. The ISS approach aims at the microscopic
description of the dipole strength function of the many transitional nuclei 
ranging between the regions of spherical and well deformed shapes, which
execute large shape fluctuations.
The special focus of our calculations is the behavior of dipole strength in the
energy range of few  MeV around the particle-emission threshold. These energies
are most important for the mentioned applications. It is also the range where
there are great gaps in theoretical and experimental knowledge about the
dipole strength function.  ISS was suggested in Ref.~\cite{zha09}, where
it was applied to the Mo, Zr, and Sr isotopes. Further it was  applied to $^{139}$La
\cite{Makinaga10}. 
The present work describes the model in
detail and presents additional results of systematic calculations in various regions of
the nuclear chart.
Our paper is organized as follows. In section \ref{sigabs}, basic features of the 
photo-nuclear absorption cross section are recalled. The ISS is introduced in section \ref{iss}.
Section \ref{iba} presents the calculational scheme of ISS. The version of the  Quasiparticle 
Random Phase Approximation used in this paper is laid out in section \ref{qrpa}.
Section \ref{results} contains the detailed discussion of our results.
The range of validity of our ISS approach is discussed in \ref{range}. Conclusions are drawn in section \ref{conc}

\section{The absorption cross section}
\label{sigabs}

The cross section $\sigma_{E1}(E)$ for the absorbtion of electric dipole ($E1$)
radiation at the energy $E$ by an even-mass nucleus is
 \begin{equation}
    \sigma_{E1}(E)= 4.037\, E\,
S_{E1}(E) \,,\quad\quad
 S_{E1}(E) = \frac{B(E1;E)\uparrow}{dE},
\label{sigmaE1}
\end{equation}
where the strength function $S_{E1}(E)$ is the derivative of the reduced
transition probability $B(E1;E)\uparrow$ for a transition from the 0$^+$ ground
state  to a $1^-$ excited state at energy $E$. The units in Eq.~(\ref{sigmaE1})
are $E$ in MeV, $\sigma_{E1}$ in mb and $B(E1)$ in e$^2$ fm$^2$. For magnetic
dipole ($M1$) radiation one has
\begin{equation}
\sigma_{M1}(E)= 0.0452\, E\,
S_{M1}(E) \,,\quad\quad
S_{M1}(E) = \frac{B(M1;E)\uparrow}{dE},
\label{sigmaM1}
\end{equation}
where the strength function $S_{M1}(E)$ is the derivative of the reduced
transition probability $B(M1;E)\uparrow$ for a transition from the 0$^+$ ground
state to a $1^+$ excited state at energy $E$. The units in Eq.~(\ref{sigmaM1})
are $E$ in MeV, $\sigma_{M1}$ in mb and $B(M1)$ in $\mu_N^2$. The absorption
cross section enters directly the total cross section for ($\gamma$,{\it particle})
reactions. For the ({\it particle},$\gamma$) reactions, the dipole strength function
determines the $\gamma$-cascade depopulating excited states. We present only results for the absorption
cross sections,  because the strength function  can be
easily obtained from the absorption cross sections by means of
Eqs.~(\ref{sigmaE1}, \ref{sigmaM1}).

The $E1$ part of the absorption cross section dominates in the energy range
above 6 MeV, where the $M1$ part amounts to only a few percent of the total
cross section (see Sec.~\ref{qrpa}). Thus, in reviewing previous work we focus
on the electric part. The prominent structure of the cross section
$\sigma_{E1}(E)$ is the Giant Dipole Resonance (GDR). 
It can be approximated by a Lorentzian
\cite{Dietrich88,sigmaBMII}
\begin{equation}
\label{sigexp}
\sigma(E,\Gamma) = \sigma_R\,\frac{E}{E_R}
\frac{(\Gamma E)^2}{(E^2-E_R^2)^2+(E\Gamma)^2} \quad\quad
\end{equation}
and is determined by three parameters, the resonance energy $E_R$, the maximum
height $\sigma_R$, and the width $\Gamma$. This expression represents the
amount of radiation absorbed by a classical damped dipole oscillator, where
$\Gamma/\hbar$ is the dissipation rate of the vibrational energy
which determines the width of the resonance.  Concerning the
low-energy tail, the question arises if and in which way $\Gamma$ depends on
the energy. The authors of Ref.~\cite{Kadmenskii83} suggest $\Gamma \propto E_x^2$ based on
the theory of Fermi liquids, where $E_x$ is the photon energy. Other authors consider the width as constant.    
For axially deformed nuclei, the GDR splits into two peaks
\cite{Okomoto,Danos}, for triaxial deformation into three peaks, where each 
of the modes can be described by a Lorentzian \cite{jun08}.  The different resonance
frequencies reflect the different wavelengths along the principal axes of the deformed nucleus
\cite{sigmaBMII}. The  Hauser-Feshbach codes, used to calculate reaction rates
for various applications, like such as  the one in Ref.~\cite{Raus00},
traditionally employ some version of the two-Lorentzian model \cite{Thie83}. The
deformation parameters of the assumed axial shape are taken either from
experimental $B(E2)$ values or from calculations by means of the Micro-Macro (MM)
Method \cite{Moeller1,Moeller2}. 

In Ref.~\cite{jun08} the experimental
absorption cross section was analyzed in terms of a model of three
individual Lorentzians with relative resonance energies related to the nuclear
deformation according to hydrodynamics  (see,
e.g. Ref.~\cite{sigmaBMII}) and $\Gamma \propto E_{R}^{1.6}$, where $E_R$ is the energy of
the resonance. Recently, cross
sections in the GDR region were combined with novel photon-scattering data
obtained from experiments at the ELBE accelerator
\cite{88Sr-data,90Zr-data,89Y-data,98100Mo-data,Mo-data}, which provided for the
first time cross sections from the low-energy region up to the GDR region. The
new data suggest  that the smooth Lorentzian extrapolation to energies far
below the peak region of the GDR provides a rough estimate for the average
trend of the cross sections, but is not capable of describing resonance
phenomena observed in this energy region. Although the experimental information
about the cross section below the particle emission threshold is still sparse
there is evidence for pronounced fluctuations and resonance-like structures
that are above the Lorentzian  \cite{88Sr-data,90Zr-data,89Y-data}.
Furthermore, below 5 MeV the cross section changes more and more in to a
discrete spectrum of individual 1$^\pm$ states. This energy region becomes
important in reactions involving neutron-rich nuclei. The dipole strength
function that determines the $\gamma$-cascade in ({\it particle},$\gamma$) reactions
may also belong to this energy region if the $Q$ value is low.  

\begin{figure}
\includegraphics[height=7cm]{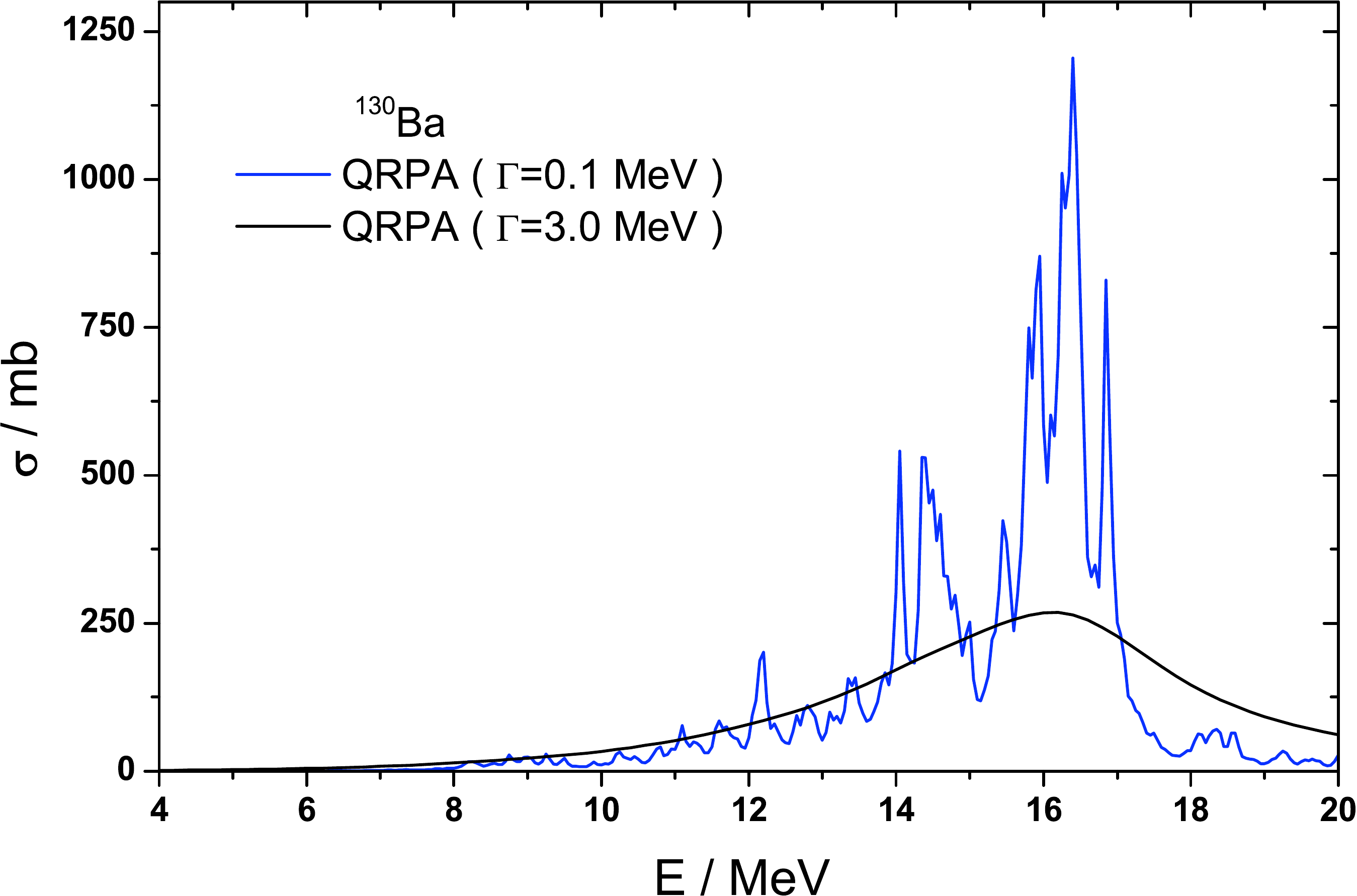}
\caption{(Color online) Absorption cross section of $^{130}$Ba calculated in QRPA with 
the equilibrium deformation $\beta=0.171$ (cf. Tab.~\ref{equilibriumDef}). Solid black
curve: width $\Gamma$ = 3 MeV, thin blue curve: width $\Gamma$ = 0.1 MeV.}
\label{QRPAspectrum}
\end{figure}

A more microscopic approach is needed to account for these shortcomings.
Moreover, a microscopically founded description of the dipole strength function
is a longstanding challenge of its own to nuclear theory. The traditional
approach is to start from the Quasiparticle Random Phase Approximation (QRPA)
 \cite{Rowe}, which extends the mean field
description of the nucleus by taking into account the quasi-bosonic part of the
residual interaction, which generates a coupling between the elementary
two-quasiparticle  excitations in the selfconsistent mean field potential
\cite{Ring}. After the QRPA dispersion relation is solved one obtains a
discrete series of states that are superpositions of the two-qpasiparicle excitations.
These states describe how the cross section of the collective dipole vibration
is spread over the two-quasiparticle excitations. This mechanism is called Landau
fragmentation (or Landau damping in the case of a dense two-quasiparticle spectrum). It
accounts for a part of the width of the GDR. 

For example in Fig. \ref{QRPAspectrum}, we show the absorption cross section
for the axially deformed nucleus $^{130}$Ba. The discrete energies of the QRPA solutions
 are folded
with a Breit-Wigner  distribution with a width of $\Gamma$ = 0.1MeV (see Eq.~(\ref{eqn:sfn}) below). 
The strong fluctuations of the cross section reflect the individual
structure of the two-quasiparticle  excitations that contribute to the GDR. The nucleus is
prolate. The two main peaks correspond to the vibrations along the
long axis ($K = 0$, low) and along the short axis ($|K|$ = 1, high). The Landau
fragmentation generates a cross section $\sigma$ at low energy, which, as
discussed in the following, is of the order of the observed one.
However, there
must be strong additional couplings of the QRPA "doorway" states to more
complicated excitations, which smooth out the fluctuations completely. 

In standard QRPA calculations these couplings are taken into account in a
phenomenological way by folding the QRPA solutions with a Breit-Wigner or
Lorentz function
analogously to the three solutions of the dipole oscillator 
just mentioned. We prefer for simplicity the Breit-Wigner
distribution which is practically equivalent to the Lorentz function used in
Eq. (\ref{sigexp}). The Breit-Wigner distribution is given by 
\begin{eqnarray} 
\label{eqn:sfn}
\sigma_{_{QRPA}}(E,\Gamma) = 
\sum_\nu{\sigma_\nu\,\frac{\Gamma/2\pi}{(E_\nu-E)^2+\Gamma^2/4}},
\end{eqnarray}
where $\sigma_\nu$ is the cross section of the QRPA solution at the energy
$E_\nu$. The cross section takes on the smooth shape seen in experiments, if
folded with a large width $\Gamma$ = 3 MeV. In the considered case the 
relation $E_R(K=1) - E_R(K=0) < \Gamma$ holds, and the two peaks merge into a
broad peak, the large width of which reflects the deformation. QRPA
calculations for static spherical or deformed shapes (see, e.g.
Ref.~\cite{sol92,skyrme-separ2,RMF-QRPA}) use such a smoothing procedure to
account phenomenologically for the neglected coupling to more complex
configurations. Accordingly, in Refs.~\cite{GorielySF,Ripl2} strength functions
for applications in Hauser-Feshbach codes were calculated throughout the
nuclear chart. The authors start from  the Skyrme Hartree-Fock-Bogoliubov (HFB)
mean field, which provides the deformation parameters.  However, they carry out
the QRPA calculations for spherical shapes only and include the splitting
caused by deformation in a phenomenological way. Hence, QRPA describes the
parts of the GDR width that originate from Landau fragmentation and static
deformation. 

The microscopic origin of the remaining part of the spreading of
the GDR has 
been reviewed, for example, in Ref.~\cite{Bertsch83,Wambach88}. The escape 
width $\Gamma_{\uparrow}$ accounts for the emission of particles from the QRPA
states. It is only of importance for light nuclei or nuclei near the drip
lines. The spreading width $\Gamma_{\downarrow}$ describes the coupling to 
more complex configurations. It is suggested that the coupling to correlated
four-quasiparticle states should contribute to $\Gamma_\downarrow$, whereas the coupling to six-quasiparticle
and higher-order quasiparticle excitations will only wash out some residual fluctuations
of the strength function. For spherical nuclei, the coupling to the four-quasiparticle configurations has been
taken into account in the framework of the 
quasiparticle-phonon models, such as QPM \cite{sol92,QPCT,Tsoneva08},
QRPA-PC \cite{QRPA-PC}, QTBA \cite{QTBA1,QTBA2,QTBA3} and RQTBA
\cite{Litv09,Lit09}. The resulting strength functions reproduce the spreading
of the GDR. Although a generalization to nuclei with static deformation is
possible \cite{sol92}, calculations have been restricted to spherical nuclei so
far because of the substantial increase of the numerical work. However,
principal problems arise in transitional nuclei that undergo large-amplitude
shape fluctuations.

\section{The ISS model}
\label{iss}

In the following we suggest an alternative approach. We explicitely describe
the coupling of the dipole vibration to the two-quasiparticle excitations by means 
of the QRPA
for deformed shapes. Out of the coupling of these QRPA doorway states to the
more complex configurations we only take into account the low-energy
collective quadrupole excitations, which represent the softest mode that
couples most strongly to the dipole mode. The quadrupole mode is described by
a model that allows for large amplitude motion, i.e. one that is suited for
transitional nuclei, which are the main object of our work. The typical
frequencies $\hbar \omega(2^+)$ of {\em collective} quadrupole excitations are
smaller than 1 MeV, which means about a factor of 10 less than the energies
$\hbar \omega(1^-)$ of the dipole excitations. Because the quadrupole motion
is much slower than the dipole one we use the adiabatic approximation: By
means of QRPA, we calculate the dipole absorption cross section 
$\sigma_{E1,M1}(E,\beta_n,\gamma_n)$ for a set of instantaneous deformation
parameters $(\beta_n,\gamma_n)$ of the mean field. We determine the
probability $P(\beta_n,\gamma_n)$ of each shape being present in the ground
state and obtain the total cross section as the incoherent sum of the
instantaneous ones,
\begin{equation}\label{adiabat}
\sigma_{ISS}(E)=\sum_nP(\beta_n,\gamma_n)\left[\sigma_{E1}(E,\beta_n,\gamma_n)
               +\sigma_{M1}(E,\beta_n,\gamma_n)\right].
\end{equation}
The zero-point motion in the ground state with respect to the collective
quadrupole modes is represented by the set of instantaneous shapes
$(\beta_n,\gamma_n)$ and their probabilities $P(\beta_n,\gamma_n)$. 
As a first step, the ''dynamical'' ground state  is constructed 
in the framework of the Interacting Boson Approximation (IBA)  \cite{IBA1}.  IBA is not a compulsory choice.
Other approaches that describe large-amplitude quadrupole motion could be used as well.
However,  IBA treats the large amplitude quadrupole motion in an efficient way and has
proven to be successful for a systematic description of transitional
nuclei. In practice it is very easy to handle because it has a minimum set of
free parameters. Moreover, it allows us to generate the discrete set of  
instantaneous shapes $(\beta_n,\gamma_n)$ and the probabilities
$P(\beta_n,\gamma_n)$ in a simple way. In the next step, a series of QRPA
calculations is performed, where $(\beta_n,\gamma_n)$ defines the shape of the
Woods-Saxon (WS) potential in the QRPA Hamiltonian. Then the total cross
section is obtained as the incoherent sum Eq.~(\ref{adiabat}) of the
respective cross sections $\sigma_{E1}(E,\beta_n,\gamma_n)$ and
$\sigma_{M1}(E,\beta_n,\gamma_n)$ multiplied by the probabilities
$P(\beta_n,\gamma_n)$.  Finally, the coupling to the more complex 
configurations is taken into account by folding $\sigma_{ISS}(E)$ with the
Breit-Wigner function
\begin{eqnarray} 
\label{eqn:cd}
\sigma_{_{ISS+CD}}(E) = \int dE' {\sigma_{ISS}(E')\frac{\Gamma(E')/
                        2\pi}{(E-E')^2+\Gamma(E')^2/4}}.
\end{eqnarray}
The width is chosen to depend quadratically on the photon energy $E$ as
expected for collisional damping (CD), i.e. $\Gamma(E)=\alpha E^2$
\cite{Bertsch83,Kadmenskii83,Wambach88}. We will refer to this
phenomenological correction as "collisional damping" (ISS+CD) although it
comprises all kinds of couplings that are not explicitly taken into account.
The value $\alpha = 0.0111$ used in our calculations corresponds to
$\Gamma(E) = 2.5$ MeV at $E =$ 15 MeV. The relation
$\hbar \omega(2^+)/\hbar \omega(1^-)\ll 1$ is taken as a justification for
neglecting possible phase correlations between the different shapes
$(\beta_n,\gamma_n)$. In other words, we assume that the deformation does not
change during the excitation of the nucleus by the absorbed photons, that is, 
the photon "sees" the shape of the nucleus that absorbs it. Thus, the photon
current on a target samples the  various  instantaneous shapes of the nuclei in the
ground state. Accordingly we suggest the name Instantaneous Shape Sampling 
(ISS) QRPA for the approach. In the following the different steps of the
ISS-QRPA approach are explained in more detail, results of calculations for
a selection of nuclei 
are presented, and
a more sophisticated 
justification of the ISS procedure is given.

\section{The calculation of the instantaneous shapes and probabilities within IBA}
\label{iba}

\begin{figure}
\includegraphics[height=7cm]{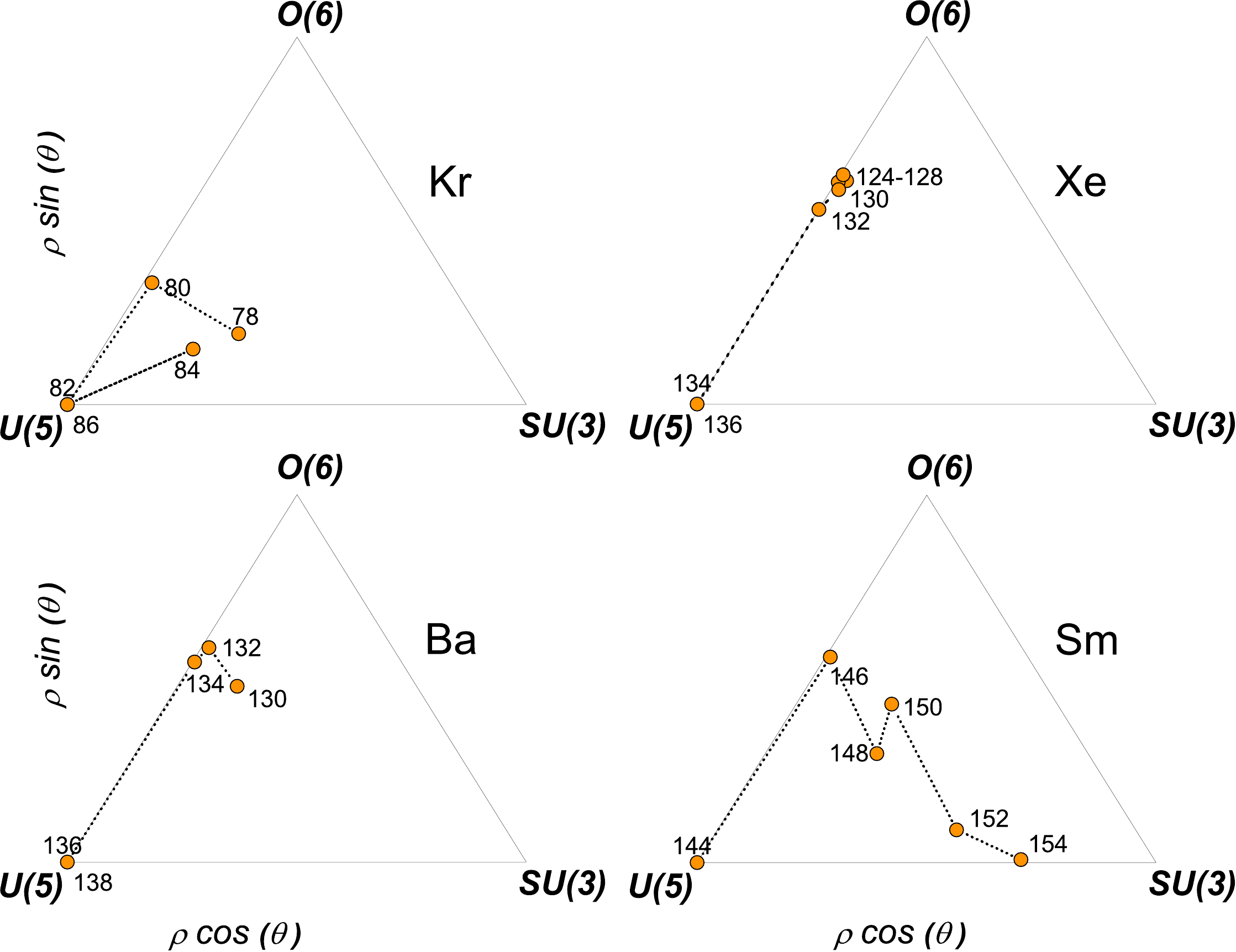}
\caption{(Color online) IBA symmetry triangles for the even-$A$ isotopic chains $^{78-86}$Kr,
$^{124-136}$Xe, $ ^{130-138}$Ba and $^{144-154}$Sm.}
\label{triangles}
\end{figure}

The family of collective quadrupole states in transitional nuclei is described
by the simplified IBA Hamiltonian suggested in
Refs.~\cite{McCutchan04,Casten-IBA}
\begin{eqnarray}
\label{H-IBA}
H_{_{IBA}}=c\,[\,(1-\zeta ) \, \hat n_d - \frac{\zeta}{4N_b}\, Q\cdot Q \,],
\\
\label{Q-IBA}
Q_\mu = s^\dagger d_\mu + d_\mu^\dagger s 
      + \chi \,[ d^\dagger\otimes d \,]\,_{2\mu}.
\end{eqnarray}
The operators $s^\dagger, s$ and $d_\mu^\dagger, d_\mu$ denote the $l=0$
and $l=2$ boson operators, respectively, of the IBA-1 model \cite{IBA1}, and
$\hat n_d$ is the number operator of d-bosons. The factor $c$ sets the energy
scale and has no influence on the structure of the states. 
The isoscalar quadrupole operator $Q_\mu$ appearing in the Hamiltonian is also
used to define the electric quadrupole (E2) transition operator
 \begin{equation}\label{eqn:TIBM}
T_\mu(E2) = e_b Q_\mu,
\end{equation}
where $e_b$ represents the boson effective charge
 in units e~b.
 In IBA applications the
total number $N_b$ of $s$ and $d$ bosons is usually chosen to be half of the
number of valence particles or holes relative to the nearest closed shells in
the considered nucleus. In contrast, we fix the values to $N_b=10$. The reason
is that $N_b$ determines the number of instantaneous shapes to be sampled.
The small values of $N_b$ near closed shells would not allow us to represent
the fluctuations of the shape of these nuclei. Fitting the IBA parameters, we
found that we could reproduce the low-energy states with the same accuracy as
for the traditional IBA choice of $N_b$. The range of the essential parameters 
$\zeta$ and $\chi$ in (\ref{H-IBA}, \ref{Q-IBA}) is restricted to the intervals [\,0,1\,] and
[\,0,\,\,-\,$\sqrt{7/2}$\,], repectively. Within this range, which defines the
so-called symmetry triangle shown in Fig.~\ref{triangles}, the whole variety of transitional
structures  between the vibrational, rotational and $\gamma$-independent nuclei
is included. For a given nucleus the values of $\zeta$ and $\chi$ are searched
by a fit to the experimental energy ratios $E(4_1)/E(2_1)$, 
$E(0_2)/E(2_1)$, and $E(2_2)/E(2_1)$. The resulting IBA
parameters values $(\zeta,\chi)$ for the nuclides considered in this paper are 
collected in Tab.~\ref{tab:IBApar}.  Fig.~\ref{triangles} places the values for the Kr, Xe, Ba and Sm
isotopes   into the symmetry triangles, where  the definition
of the polar coordinates ($\rho, \theta$) according to 
Ref.~\cite{McCutchan04} is applied. Such contour lines within the symmetry
triangle may help to extrapolate to nuclei with less well 
known excitation spectra. 

\begin{table}
\caption{\label{tab:IBApar}Optimal IBA parameters $\zeta$, $\chi$ and the boson
                           charge $e_b$ for the considered nuclides.} 
\begin{ruledtabular}
\begin{tabular}{rrccc}
$A$     &$N$ &$\zeta$&$\chi$& $e_b$ in e~b  \\
\hline
Kr: ~78 & 42 & 0.47 & -0.79 &  0.088 \\
     80 & 44 & 0.35 & -0.06 & 0.075   \\
     82 & 46 & 0.0  & -1.2  & 0.067  \\
     84 & 48 & 0.35 & -0.76 & 0.044  \\
     86 & 50 & 0.0  & -1.2  & 0.049 \\
        &    &      &       &        \\
Xe: 124 & 70 & 0.63 & -0.04 & 0.094 \\
    126 & 72 & 0.61 & -0.01 & 0.086  \\
    128 & 74 & 0.63 & -0.01 & 0.083  \\
    130 & 76 & 0.60 & -0.03 & 0.079  \\
    132 & 78 & 0.53 & -0.0  & 0.074  \\
    134 & 80 & 0.0  & -1.20 & 0.082  \\
    136 & 82 & 0.0  & -1.20 & 0.085  \\
        &    &      &       &        \\
Ba: 130 & 74 & 0.61 & -0.26 & 0.105  \\
    132 & 76 & 0.60 & -0.03 & 0.091  \\
    134 & 78 & 0.55 & -0.01 & 0.084  \\
    136 & 80 & 0.0  & -1.20 & 0.091  \\
    138 & 82 & 0.0  & -1.20 & 0.068  \\
        &    &      &       &        \\
Sm: 144 & 82 & 0.0  & -1.20 & 0.072  \\
    146 & 84 & 0.57 & -0.02 & 0.071  \\
    148 & 86 & 0.54 & -0.59 & 0.087  \\
    150 & 88 & 0.64 & -0.41 & 0.108  \\
    152 & 90 & 0.62 & -1.15 & 0.146  \\
    154 & 92 & 0.71 & -1.31 & 0.142  \\
\end{tabular}
\end{ruledtabular}
\end{table}

The next step is the calculation of the probability distribution
$P(\beta_n,\gamma_n)$ of the deformations $\beta_n$ and $\gamma_n$ in the $0^+$
ground state of the boson Hamiltonian Eq. (\ref{H-IBA}). We follow the method
proposed in Refs.~\cite{Werner00,Tonev07}. We relate the IBA to the shape of the
Woods-Saxon potential used in QRPA by assuming that its deformation parameters
$\beta,\gamma$ are the same as in the expression for the electric quadrupole
$(E2)$ transition operator of a charged liquid drop,
\begin{equation}\label{eqn:TLD}
T_\mu(E2) = \frac{3ZeR^2}{4\pi}\beta\left[D^2_{\mu0}\cos\gamma
        + \left(D^2_{\mu2}+D^2_{\mu-2}\right)\frac{\sin\gamma}{\sqrt{2}}\right],
\end{equation}
where the $D^2_{\mu\nu}$ functions transform the quadrupole moments from the body-fixed
frame to the laboratory frame in the standard way. From Eqs.~(\ref{eqn:TIBM})
and (\ref{eqn:TLD}) it follows that the two scalar invariants constructed from
the IBA quadrupole operator $Q_\mu$ given by Eq.~(\ref{Q-IBA}) are connected
with the deformation as follows
\begin{eqnarray}
\label{q-def}
\hat q_2&=&\left[ Q \otimes Q \right]_{0} \propto \beta^2,\\
\label{q-def2}
\hat q_3&=&\left[ Q \otimes 
\left[ Q \otimes Q \right]_{2} \right]_{0} \propto \beta^3 \cos{3\gamma}. 
\end{eqnarray} 

A set of localized states  $|n\rangle$ is generated by  diagonalizing
$\hat q_2$ and $\hat q_3$ within the basis of \mbox{$N_b=10$} of s-d-boson states. 
Because the scalars $\hat q_2$ and $\hat q_3$ commute, they can be
simultaneously diagonalized. As we are interested only in the probability
distribution $P(\beta_n,\gamma_n)$ of the IBA ground state $|0^+_1\rangle$,
the diagonalization is restricted to the set of 0$^+$ basis states within the
boson space of maximal $N_b=10$ d-bosons. The eigenvalues $q_{2,n}$ and
$q_{3,n}$ provide the values of the deformation parameters ($\beta_n$,$\gamma_n$)
that are assigned to each localized state $|n\rangle$  
by the relations 
\begin{equation}
\label{scalarsx}
\beta_n^2 = \sqrt{5}\left(\frac{4\pi e_b}{3ZeR^2}\right)^2q_{2,n}~, ~~~~~~
\cos{3\gamma_n} = \sqrt{\frac{7}{2\sqrt{5}}}\frac{q_{3,n}}{(q_{2,n})^{3/2}}
\end{equation} 
where $e_b$ is the effective boson charge and $R=1.2A^{1/3}$\,fm is the nuclear charge
radius (cf.  Eqs. (\ref{eqn:TIBM}) and (\ref{eqn:TLD})). The eigenstates
$|n\rangle$ of the operators (\ref{q-def}, \ref{q-def2}) are identified with
''instantaneous'' mean field states with the deformation parameters
$(\beta_n,\gamma_n)$. As usual, the boson charge $e_b$  is adjusted  to measured
$B(E2,2^+_1\rightarrow 0^+_1)$ values \cite{IBA1}, which are taken from the compilation 
\cite{raman}. The
probabilities $P(\beta_n,\gamma_n)$ are finally obtained by projecting the 
eigenstates $|n\rangle$ onto the IBA ground $|0^+\rangle$, i.e.
\begin{equation}
\label{Pn}
P(\beta_n,\gamma_n)=|\langle 0^+_1|n\rangle|^2.
\end{equation}  
Our procedure assumes that the instantaneous charge
density and the instantaneous mean field have the same deformation.

\begin{figure}
\includegraphics[height=7cm]{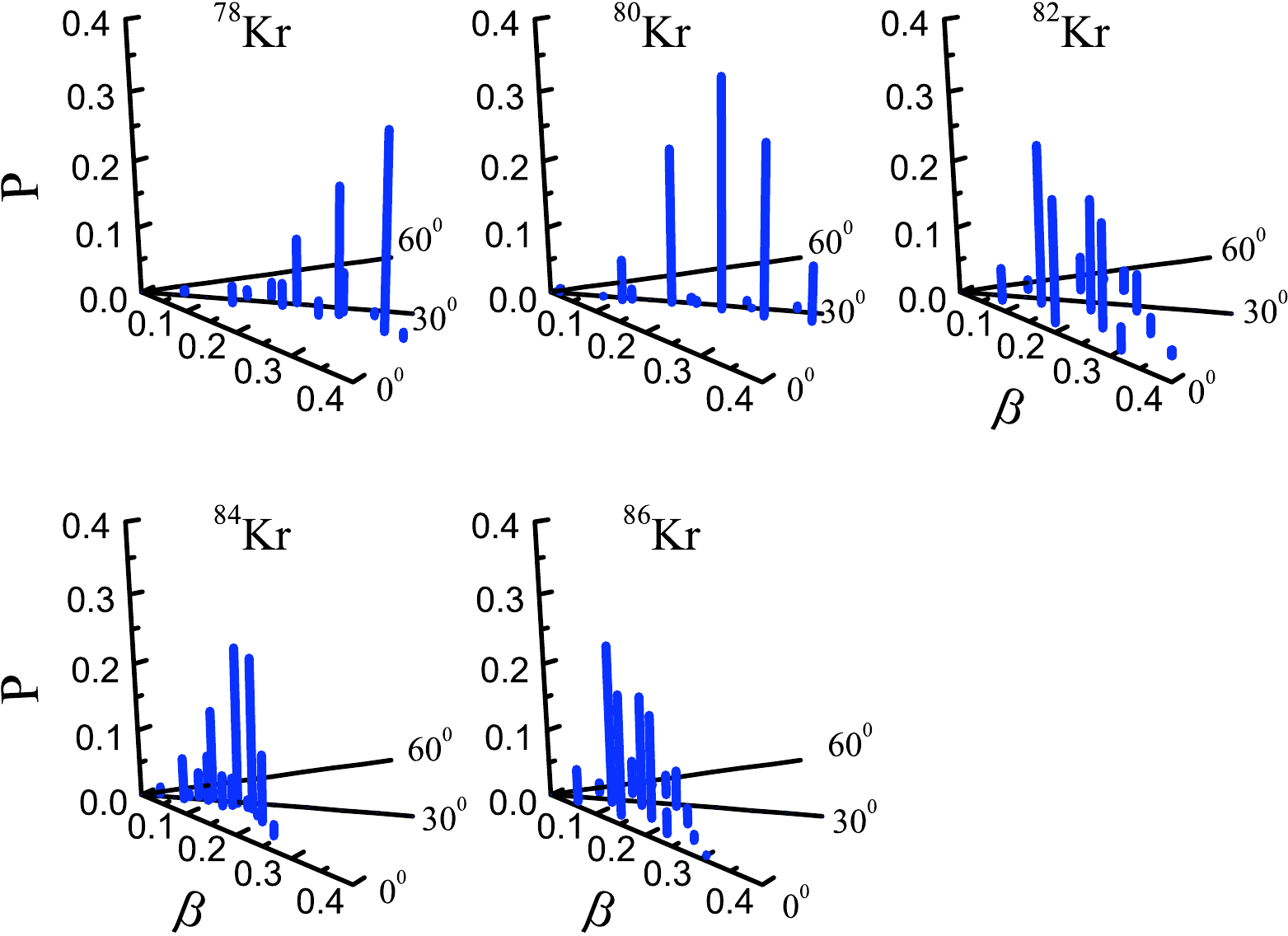}
\caption{\label{KrDistr}(Color online) Probability distributions for $^{78-86}$Kr. The
instantaneous nuclear shapes over the $\beta-\gamma$ plane were found by means
of the IBA with the parameters given in Table~\ref{tab:IBApar}.}
\end{figure}

As an example, we discuss  the IBA part of the ISS calculations for the even-$A$
chain $^{78-86}$Kr. In the upper part of Table~\ref{tab:IBApar} the  IBA parameters 
$(\zeta,\chi)$ are given, and in Fig.~\ref{KrDistr} the resulting instantaneous
deformations $(\beta_n,\gamma_n)$ and the probabilities $P_n$  are
displayed. Note that $(\beta,\gamma)$ are curvilinear coordinates with the volume
element $d\beta\beta^4d\gamma \sin3\gamma$. These geometric factors are part
of the probabilities $P_n$. This has the consequence that $\gamma=30^o$ is favored. For
example, $^{80}$Kr has $\chi \approx 0$, which means that it tends to a
$\gamma$-instabliblity between the U(5) -- O(6) limits
(cf. Fig.~\ref{triangles}). The corresponding distribution in 
Fig.~\ref{KrDistr} looks as if the nucleus had a stable triaxial deformation,
which however only reflects the factor $\sin3\gamma$ in the volume element. 
In a similar fashion, $\beta = 0$ is suppressed. For example, $^{86}$Kr has $\zeta=0$,
which means it is a spherical  vibrator (U(5) limit). The corresponding
distribution in  Fig.~\ref{KrDistr} looks as if the nucleus had a stable deformation,
which however only reflects the factor $\beta^4$ in the volume element.  

\section{The QRPA }
\label{qrpa}

The Hamiltonian $H$ for the QRPA \cite{Ring} comprises the mean field part
$h_{mf}$ and a residual dipole-dipole interaction $v_{res}$,
\begin{eqnarray}
\label{eqn:QRPA}
H = h_{mf}^{(p)}(\beta_n,\gamma_n) + h_{mf}^{(n)}(\beta_n,\gamma_n) 
   + v_{res},\nonumber\\
\label{eqn:QRPA1}
h_{mf}^{(\tau)}(\beta_n,\gamma_n) = h_{WS}^{(\tau)}(\beta_n,\gamma_n)
+ \Delta^{(\tau)}(P^{+(\tau)} + P^{(\tau)}) - \lambda^{(\tau)} N^{(\tau)}, 
\end{eqnarray}
where the indices $\tau=p,n$ refer to  protons and neutrons, respectively.
The spatial part of the mean field $h_{WS}^{(\tau)}(\beta_n,\gamma_n)$ is the
triaxial Woods-Saxon (WS) potential (for parameters see Ref. \cite{Kaha89}).
The  deformation parameters $(\beta_n,\gamma_n)$ of the WS potential define  
the instantaneous quadrupole-deformed shape about which the nucleus executes the
isovector dipole oscillations. In order to describe the partial occupancies of
the  single-particle levels in open-shell nuclei a static monopole pair
potential is supplemented to the WS part in Eq. (\ref{eqn:QRPA1}), which is
defined  by the gap parameter $\Delta^{(\tau)}$ and the Fermi energy
$\lambda^{(\tau)}$. Therein, $N^{(\tau)}$ and $P^{(\tau)+}$ denote the particle
number and monopole pairing operators for the protons and neutrons, 
respectively.

The residual interaction consists of two terms, 
$v_{res}=v^{(E1)}_{res}+v^{(M1)}_{res}$, where the index $E1$  refers to the
negative-parity electric excitations and the index $M1$  to the positive-parity magnetic
excitations. The $E1$ and the $M1$ dipole excitations of the QRPA are
calculated separately, and their contributions to transition strengths are 
added up. Our QRPA calculations assume schematic interactions of the
dipole-dipole type. The $E1$ part is given by the electric dipole-dipole
interaction 
\begin{eqnarray}
\label{eqn:VresE1}
v^{(E1)}_{res}=
\frac{1}{2}\varkappa_{t=0}\bigl(\,\sum_{\,i=1,A}\vec x''_i\,\bigr)^{2}
+ \frac{1}{2}\varkappa_{t=1}
\bigl({\small\sum_{\,i=1,A}}\tau_i~\vec x''_i\bigr)^{2}, 
\end{eqnarray}
where here $\tau = \pm 1$ holds for neutrons and protons, respectively.
The interaction  $v^{(E1)}_{res}$ is expressed in terms of the doubly stretched
coordinates $\vec x''$ referring to the nuclear self-consistency model
\cite{sakamoto}. The term $v^{(E1)}_{res}$ is the simplest ansatz for a residual
interaction with the signature 1$^-$. The inclusion of an octupole-octupole
term has been investigated too. It is not taken into account here, because it was
found that it had 
practically no effect on the $E1$ cross section. It is worth mentioning that  
according to the investigations \cite{skyrme-separ1,skyrme-separ2}, the
isovector dipole term is the most important contribution in an expansion of a
realistic Skyrme-type interaction into separable interaction terms.  
As suggested in Ref.~\cite{Doen07}, the isoscalar $(t=0)$ interaction term is
used for removing the spurious center-of-mass motion. 
Choosing a value $\varkappa_{t=0}=~1000$ MeV~fm$^{-2}$ for the isoscalar coupling
constant ensures that the QRPA states have no spurious contributions. The
isovector strength constant $\varkappa_{t=1}$ determines the mean position of
the GDR. The $A$-dependence of the isovector strength $\varkappa_{t=1}$
is assumed to be given by the selfconsistent strength factor \cite{sakamoto} 
\begin{eqnarray}
\label{sc.const}
\varkappa_{t=1} = -\varkappa_{sc}\,\eta = -\frac{M\omega_o^2}{A}\,\eta, 
\end{eqnarray}
where $M$ is the nucleon mass and
$\hbar\omega_o = 41 A^{1/3}$ MeV. For the remaining factor $\eta$ an empirical
value $\eta\approx 3$ fm$^{-2}$ is estimated from the systematics of the
GDR peak energy. In our QRPA calculations, the $\eta$-value is kept constant
within an isotopic chain and is adjusted to the empirical peak position of the
GDR of one of the isotopes in each chain.

\begin{figure}
\includegraphics[width=8cm]{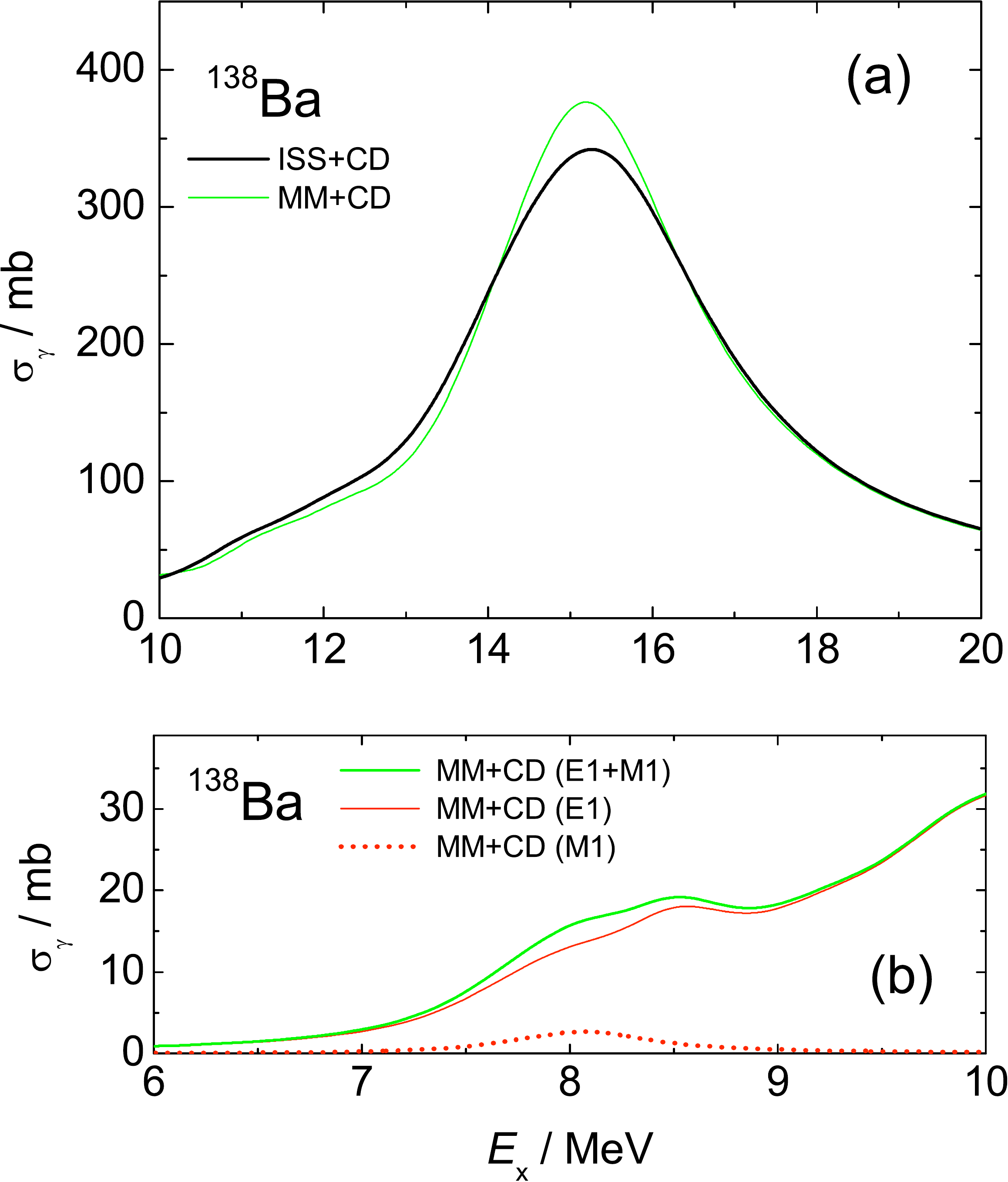}
\caption{\label{Ba138.M1+E1}(Color online) Phot-absorption cross section of $^{138}$Ba calculated with
ISS-QRPA and energy-dependent width $\Gamma(E_x) = 2.5 (E_x/15)^2$ MeV. Lower
panel (b): $E_x$ = 6 -- 10 MeV. Red solid curve: $E1$ contribution, red dotted
curve: $M1$ contribution, solid green curve: summed $E1+M1$ cross section. Upper
panel (a): $E_x$ = 10 -- 20 MeV. Black curve: QRPA with the equilibrium
deformation in Tab.~\ref{equilibriumDef}, green curve: ISS-QRPA. 
}
\end{figure}

The $M1$ part of the residual interaction $v_{res}$ is the magnetic
dipole-dipole interaction 
\begin{eqnarray}
\label{eqn:VresM1}
v^{(M1)}_{res}=
-\frac{1}{2}\sum_{t=0,1}
\kappa_j^{t}{\bf J}^{t}\cdot{\bf J}^{t}
-\frac{1}{2}\sum_{t=0,1}
\kappa_s^{t}{\bf{S}}^{t}\cdot{\bf{S}}^{t}.
\end{eqnarray}
The terms  ${\bf{J}}^t\cdot {\bf{J}}^t$  and ${\bf{S}}^t \cdot {\bf{S}}^t$ are
composed of the isoscalar ($t$ = 0) and isovector ($t$ = 1) parts of the total
angular momentum  operator $\bf{J = L + S}$ and the spin operator
${\bf{S}}$, i.e.
${\bf{J}}^{t=0,1}={\bf{J}}^{(p)}+(-1)^t {\bf{J}}^{(n)}$ and
${\bf{S}}^{t=0,1}={\bf{S}}^{(p)}+(-1)^t {\bf{S}}^{(n)}$. 
A possible quadrupole-quadrupole interaction term turned out to be 
unimportant for the $M1$ strength above 4 MeV excitation energy. The same
magnetic dipole interaction as in Eq.~(\ref{eqn:VresM1}) was used in
Ref.~\cite{PRC-M1} to describe the $M1$ properties of the Mo isotopes. In these
investigations it turned out that the $M1$ transition strength in the
interesting energy region above 5 MeV is completely dominated by spin 
vibrations that are generated  by the strong repulsive isovector spin-spin term
${\bf{S}}^{t=1}\cdot{\bf{S}}^{t=1}$ in Eq.~(\ref{eqn:VresM1}).  
The related strength parameter $\kappa_s^{t=1}$ is not well known from 
literature. Therefore, we have used the same value
$\kappa_s^{t=1}$ =~-1 MeV/$\hbar^2$ as in our previous study \cite{PRC-M1}. The isoscalar
spin-spin term ${\bf{S}}^{t=0}\cdot{\bf{S}}^{t=0}$ is neglected because the
isoscalar spin part ${\bf{S}}^{t=0}$ in the $M1$ transition operator 
is reduced by about a factor of twenty as compared to the corresponding isovector
spin part. However, a large isoscalar term  ${\bf{J}}^{t=0}\cdot{\bf{J}}^{t=0}$ 
is included by choosing $\kappa_j^{t=0}$=~1000 MeV/$\hbar^2$ in order to
eliminate effects of the spurious rotational motion. The isovector term
${\bf{J}}^{t=1}\cdot{\bf{J}}^{t=1}$ is left out because it only influences  the
scissor mode in deformed nuclei, the $M1$ strength of which appears
below 5 MeV \cite{PRC-M1}. This energy region  is not considered in the present work. 

The $M1$ strength is essentially generated by spin-flip transitions between
high-$j$ spin-orbit partners. Accordingly, it is expected to produce a summed
$B(M1)$ strength of a few $\mu_N^2$ distributed in the energy region 7 -- 9 MeV
in medium-mass nuclei (see, e.g. \cite{M1-Frekers}). This corresponds to a cross
section contribution of a few millibarn. In some cases this value can reach up
to 20 \%  of the corresponding $E1$ cross section in the low-energy region.
 As an example, we present the QRPA result for the nuclide $^{138}$Ba
in Fig.~\ref{Ba138.M1+E1}, showing the $E1$ and $M1$ contributions separately.
Because the measurement of the parity is a quite
challenging task there are hitherto only a few experiments  that identify the $M1$ part of
the dipole strength. For this reason
we do not display the relatively small $M1$
contribution to the dipole cross section separately.

Carrying out the QRPA begins with the diagonalization of the deformed WS
potential of the Hamiltonian (\ref{eqn:QRPA}) where we use an oscillator basis
with the shells $N = 0 - 8$. The pair field in Eq.~(\ref{eqn:QRPA1}) is
included in BCS approximation, which transforms the creation and annihilation
operators $c^+_k$ and $c_{k}$ of the WS levels $k$ to the quasiparticle
operators
\begin{eqnarray}
\label{eqn:qp}
a^+_k = u_k c^+_k + v_k c_{\bar k}.
\end{eqnarray}
Here, $u_k$ and $v_k$ are the usual BCS amplitudes, and   
$\bar k$  labels the time-conjugate WS levels. In terms of the 
quasiparticle operators the mean field part of the Hamiltonian
Eq. (\ref{eqn:QRPA}) takes the diagonal form
\begin{eqnarray}
\label{}
h_{mf}= \sum _k ~\varepsilon_k (a^+_{_k} a_{_k} + a^+_{_{\bar k}} a_{_{\bar k}}~),
\end{eqnarray}
where $ \varepsilon_k = \sqrt{(e_k-\lambda)^2+\Delta^2}$ are the quasiparticle
energies in the WS potential. The values of the pairing gaps $\Delta^{(p,n)}$
are derived from the binding energies by using a five-point formula.

The standard way of performing the QRPA consists of solving the equation of
motion
\begin{eqnarray}
\label{eqn:Hphonon}
[H,\Omega^+_\nu]_{_{QRPA}}=E_\nu\Omega^+_\nu 
\end{eqnarray}
for the phonon operators $\Omega^+_\nu$ (cf. Eq.~(\ref{eqn:phonon}) below )
 via a matrix diagonalization \cite{Rowe}. The suffix $QRPA$ in
Eq.~(\ref{eqn:Hphonon}) indicates that only the quasibosonic part of the 
residual interaction $v_{res}$ is included in the commutator. The set of
eigenvalues $E_\nu$ forms the discrete spectrum of the vibrational dipole
excitations. The eigenvectors of the Hamiltonian in Eq.~(\ref{eqn:Hphonon}) define the QRPA
amplitudes $\phi^{(\nu)}_{kk'}$ and $\psi^{(\nu)}_{kk'}$. The phonon operators, 
\begin{eqnarray}
\label{eqn:phonon}
\Omega^+_\nu = \sum_{kk'}
               {[\phi^{(\nu)}_{kk'}a^+_{k}a^+_{k'}+\psi^{(\nu)}_{kk'}a_{k'}a_{k}]},
\end{eqnarray}
create the vibrational states $\nu$ as a superposition of 
two-quasiparticle and two-quasihole  excitations. The partial cross
section $\sigma_\nu$ for a dipole excitation from the QRPA ground state
$|\rangle$ to a vibrational state $|\Omega^+_\nu\rangle$ at the energy
$E_\nu$ is 
\begin{eqnarray}
\label{sigmaEr}
\sigma_\nu(E)= f^{(\pm)} E_\nu |\langle\Omega_\nu 
~{\sf M^{(\pm)}_{dipole}}\rangle |^2~ \delta(E-E_\nu),
\end{eqnarray}
where ${\sf M^{(\pm)}_{dipole}}$ means the electric (-) or magnetic (+) dipole
transition operator. Measuring the cross section in mb
and the energies in MeV, the respective scale factors are
\mbox{$f^{(-)} = 4.037/$(e$^2$fm$^2)$} and \mbox{ $f^{(+)}=0.0452/(\mu_N^2)$} 
(cf. Eqs. (\ref{sigmaE1}, \ref{sigmaM1})). The total cross section $\sigma(E)$ is given by
summing over all the partial cross sections $\sigma_\nu$. Finally we replace 
$\delta(E-E_\nu)$ by a Breit-Wigner distribution of finite width $\Gamma$, which gives the
previous expression Eq. (\ref{eqn:sfn}) for the cross section $\sigma(E,\Gamma)$.

To circumvent the direct evaluation of the equation of motion
(\ref{eqn:Hphonon}) which involves  typically a large matrix diagonalization
with a rank of $n\approx 10^4-10^5$ we apply the strength function method
\cite{sol92}. With this method, the summation in Eq.~(\ref{eqn:sfn}) can be
written in terms of a contour integral which finally is cast in an analytical
formula for the function $\sigma_{_{QRPA}}(E,\Gamma)$ that is explicitly given
in Ref.~\cite{sol92}. 
The use of the analytical expression for $\sigma_{_{QRPA}}(E,\Gamma)$ leads to
an enormous simplification of the practical performance of the QRPA which is of
crucial importance for accounting of the variety of shapes $(\beta, \gamma)$
inherent to the ISS calculations.
The QRPA calculations are carried out with a constant width of $\Gamma= $0.1 MeV.
This width is small enough to retain all relevant structure of the cross section.
The method is also efficient for a separable interaction that contains more terms than 
the leading dipole-dipole term considered in our paper. The authors of Refs.
demonstrated that such separable interactions well approximated the 
non-separable interactions of the Skyrme type \cite{skyrme-separ1}.

\begin{figure}
\includegraphics[width=8cm]{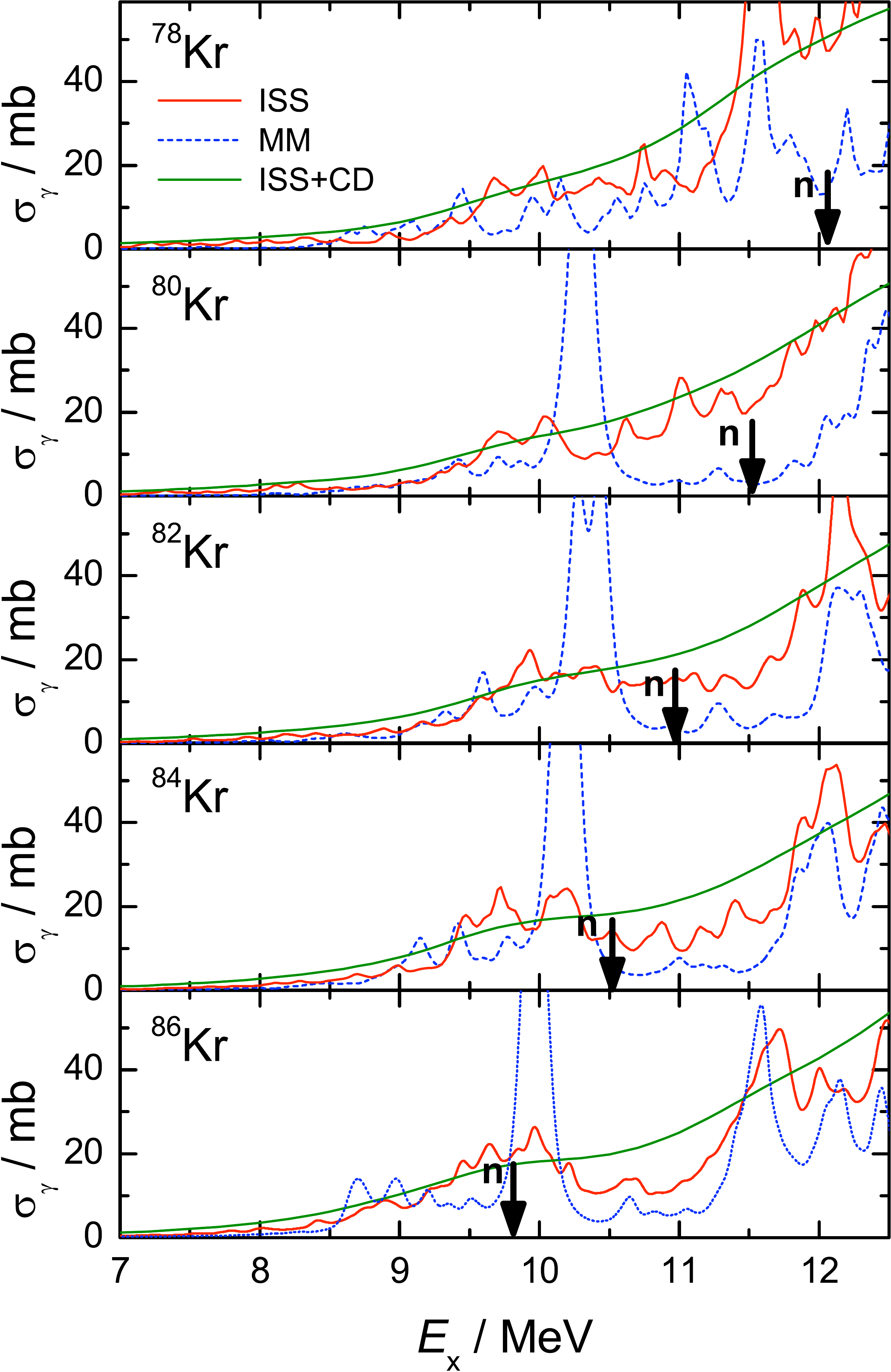}
\includegraphics[width=8cm]{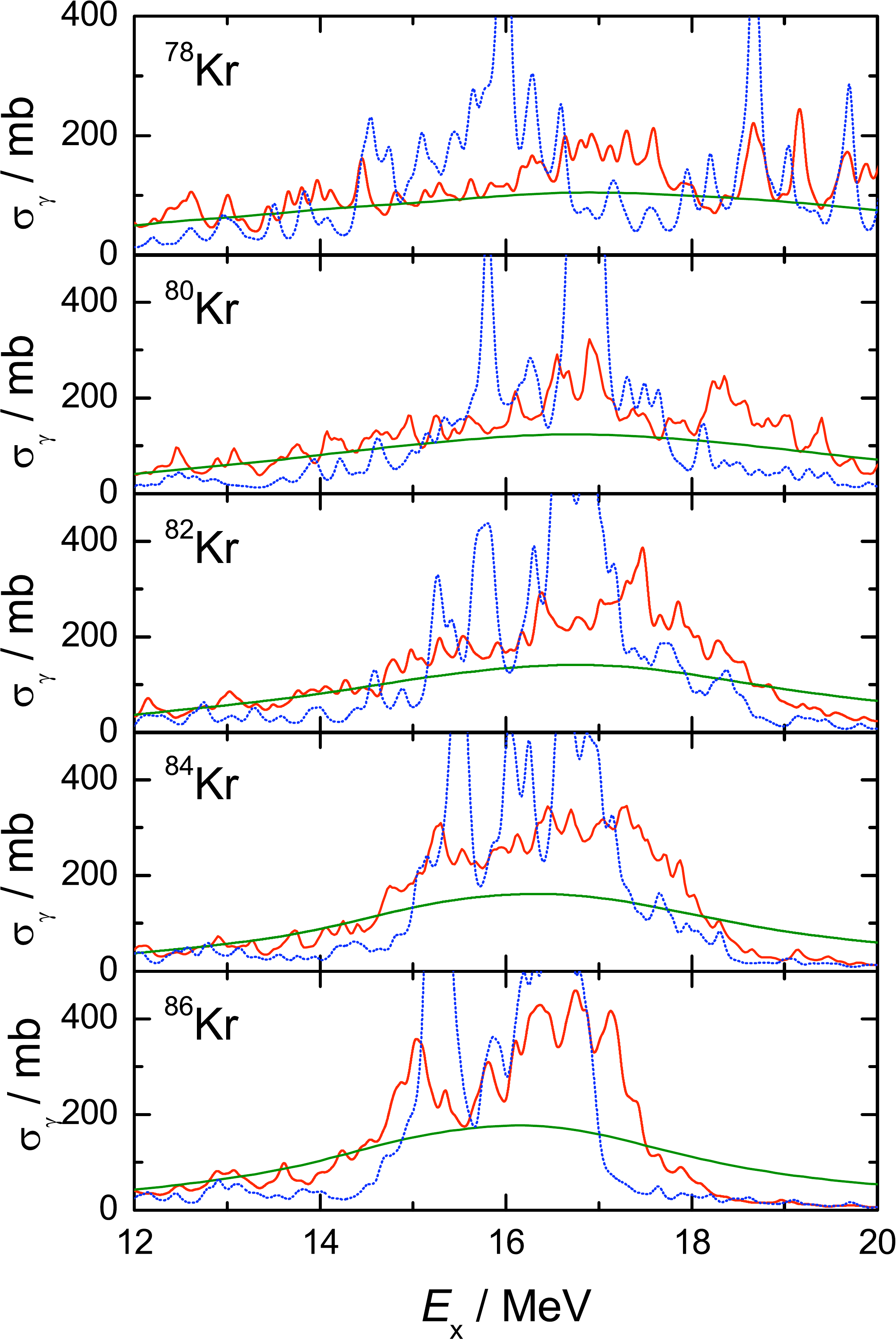}
\caption{\label{krypton}(Color online) Cross sections for the isotopes  $^{78-86}$Kr. Left
panel: $E_x$ = 7-12.5 MeV, right panel: $E_x$ = 12-20 MeV. Red curve (ISS):
calculated with ISS-QRPA ($\Gamma$=0.1 MeV). Dotted blue curve (MM): QRPA
($\Gamma$ = 0.1 MeV) with the equilibrium deformations in
Tab.~\ref{equilibriumDef}. Green curve (ISS+CD): ISS-QRPA averaged with
energy-dependent width $\Gamma(E_x) = 2.5 (E_x/15)^2$ MeV. The arrows labeled by "n"
mark the position of the neutron-emission threshold of the respective isotope.
}
\end{figure}

\begin{table}
\caption{\label{equilibriumDef}Ground-state deformation parameters from
Ref.~\cite{Moeller1,Moeller2} for the Kr, Xe, Ba, and Sm isotopes.}
\begin{ruledtabular}
\begin{tabular}
{rcr}
$A$     &$\beta$&$\gamma$\\
\hline
Kr:  78 & 0.232 &  60  \\ 
     80 & 0.062 &   0  \\
     82 & 0.071 &   0  \\
     84 & 0.062 &   0  \\
     86 & 0.053  &   0  \\
        &       &      \\
Ba: 130 & 0.171 &   0  \\
    132 & 0.158 &  20  \\
    134 & 0.132 &  30  \\
    136 & 0.0   &   0  \\
    138 & 0.0   &   0  \\
        &       &      \\
Xe: 124 & 0.208 &   0  \\ 
    126 & 0.170 &   0  \\
    128 & 0.184 &  25  \\
    130 & 0.158 &  30  \\
    132 & 0.0   &   0  \\
    134 & 0.0   &   0  \\
    136 & 0.0   &   0  \\
        &       &      \\
Sm: 144 & 0.0   &  0   \\
    146 & 0.0   &  0   \\
    148 & 0.161 &  0   \\
    150 & 0.206 &  0   \\
    152 & 0.243 &  0   \\
    154 & 0.270 &  0   \\
\end{tabular}
\end{ruledtabular}
\end{table}

\begin{figure}[b]
\includegraphics[width=8cm]{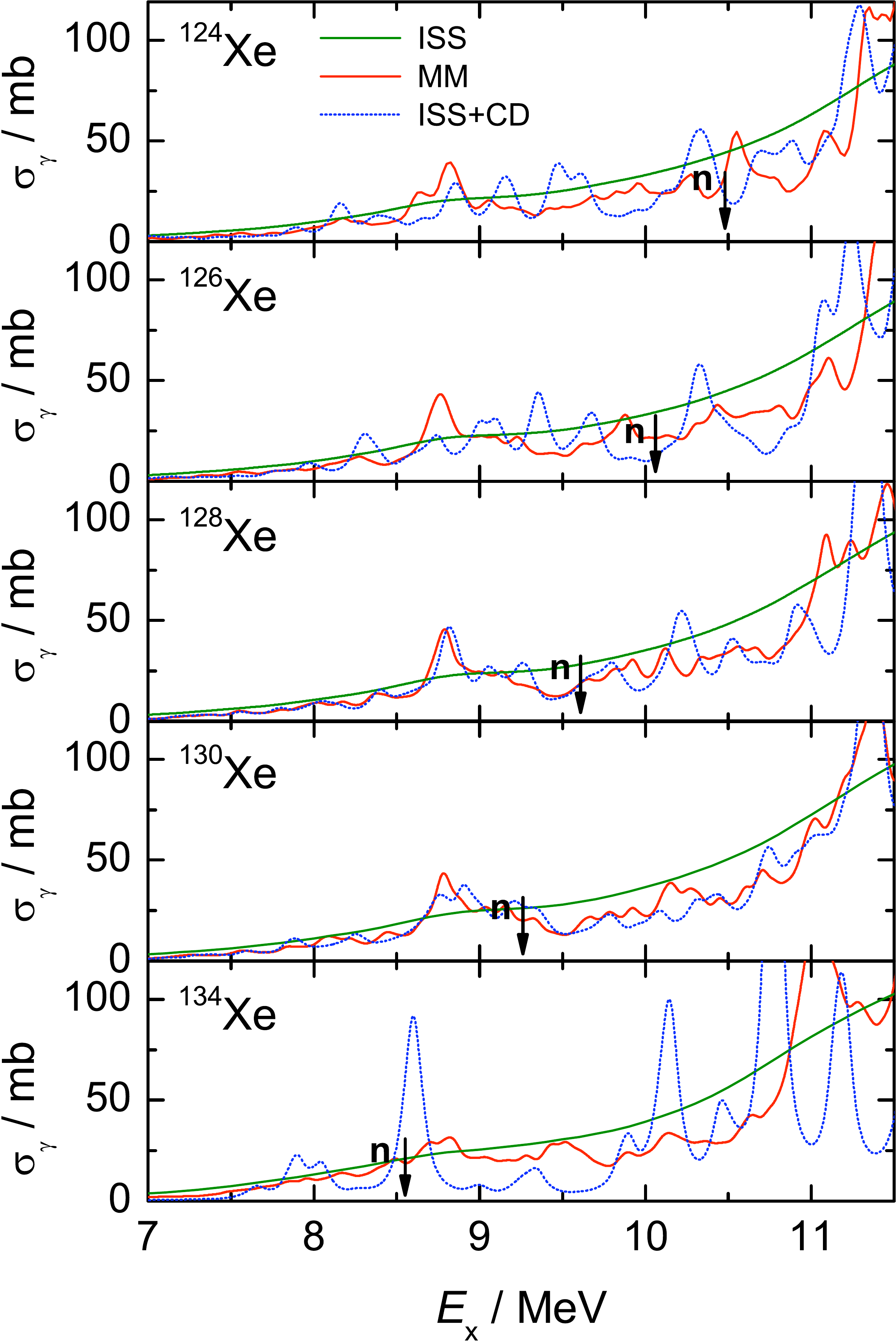}
\includegraphics[width=8cm]{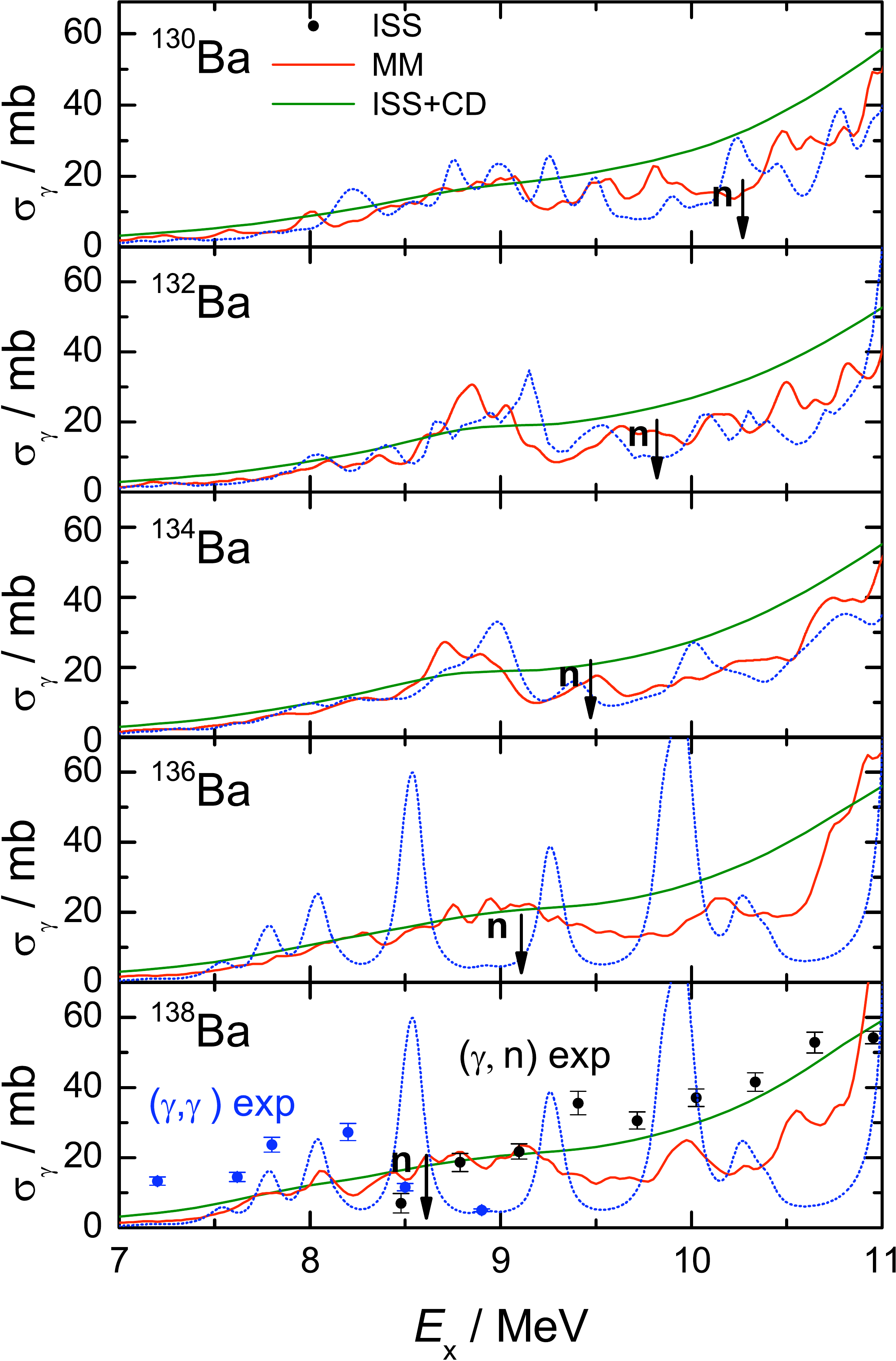}
\caption{\label{Barium}(Color online) Cross sections for the isotopes $^{124-134}$Xe 
(left panel) and $^{130-138}$Ba (right panel) in the low-energy region 7 -- 11
MeV. Notations are as in Fig.~\ref{krypton}. The arrows labeled by 
 "n" mark the
positions of the neutron-emission threshold of the isotopes. The black dots
with error bars in $^{138}$Ba display the measured $(\gamma,n)$ cross section from
Ref.\cite{Ripl2}. The blue dots below the neutron threshold are $(\gamma,\gamma')$ data 
from the recent measurement by Tonchev et al. \cite{Ba138Tonchev}.
}
\end{figure}

\begin{figure}
\includegraphics[width=8cm]{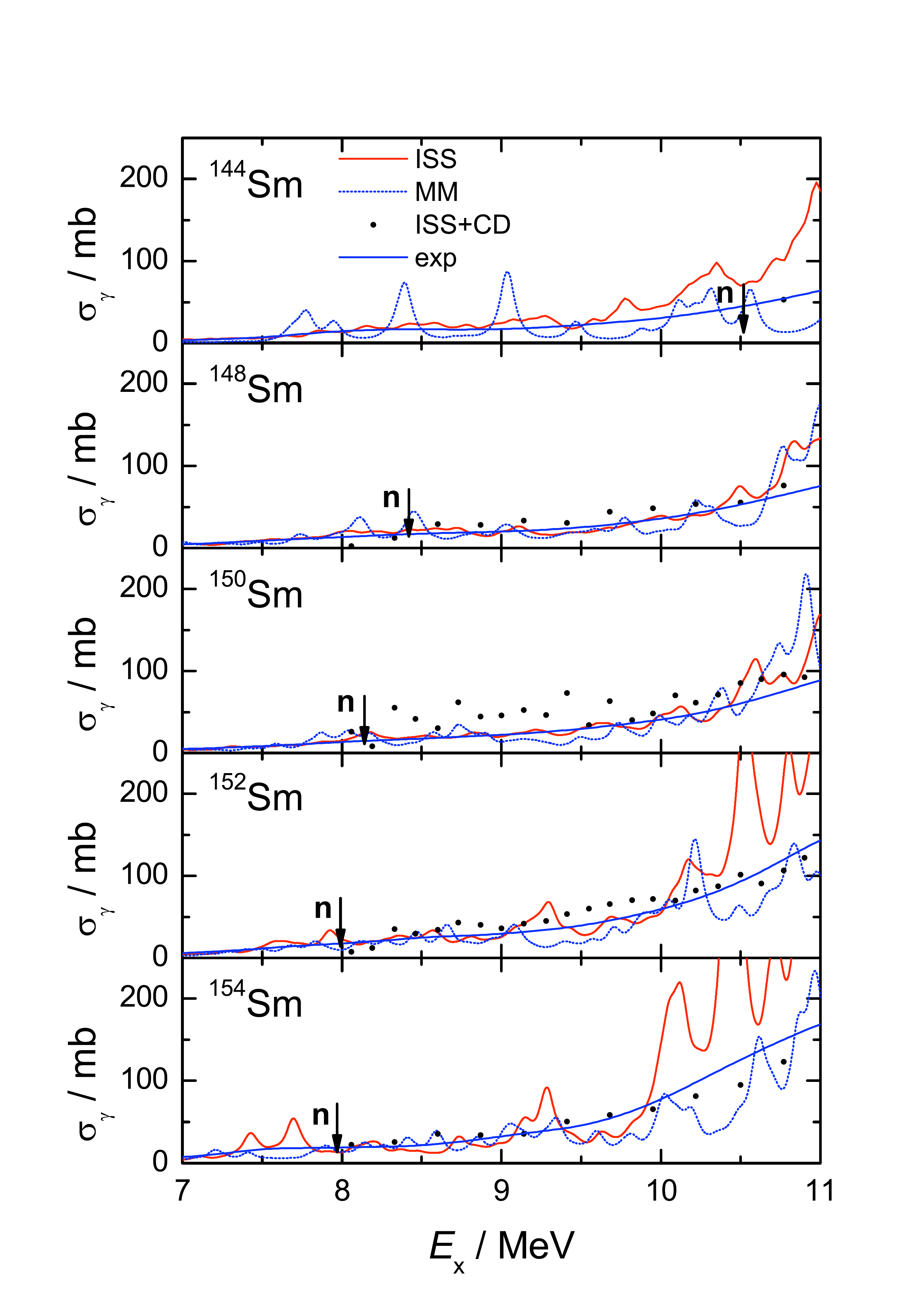}
\includegraphics[width=8cm]{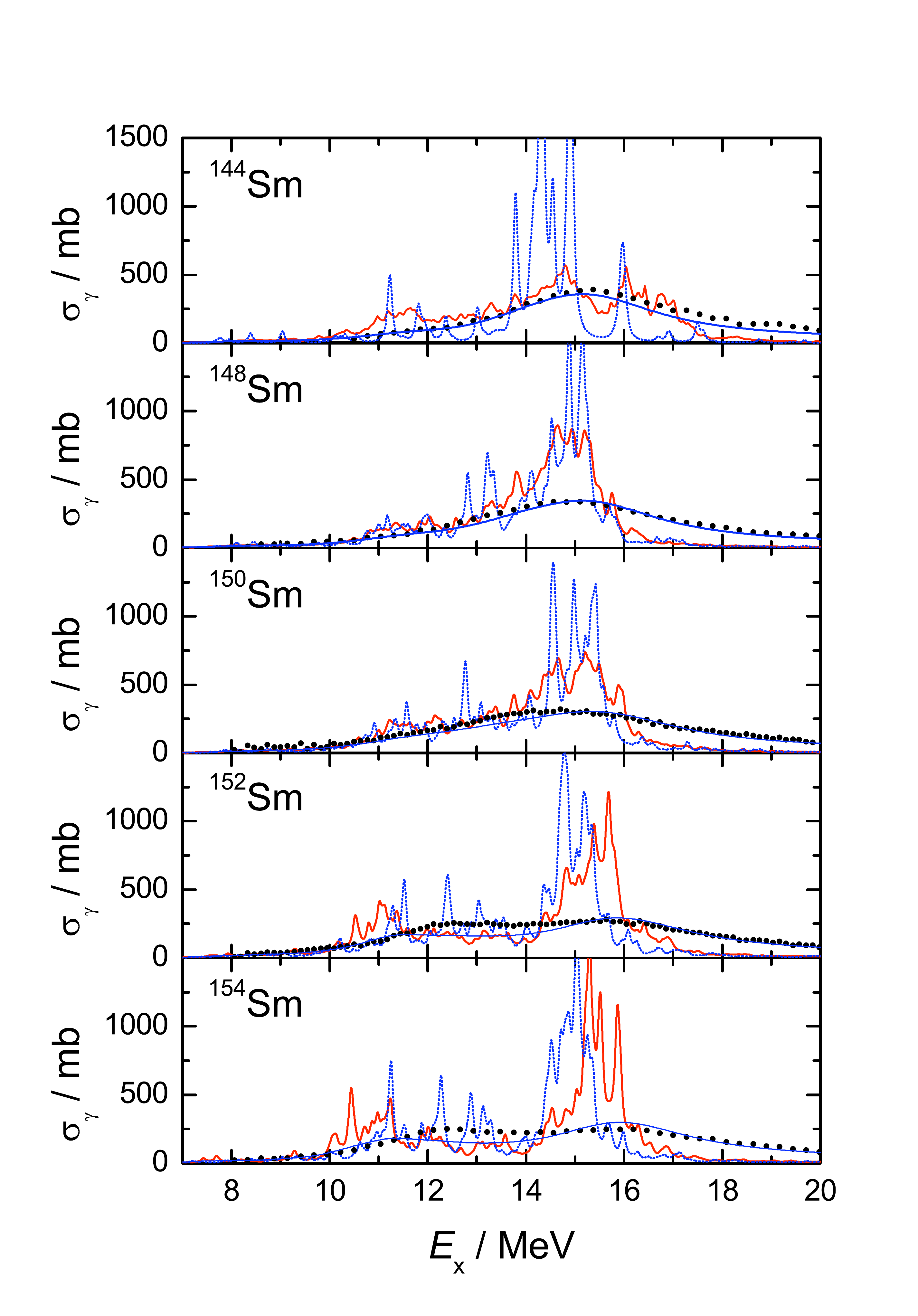}
\caption{\label{samarium}(Color online) Cross sections for the isotopes $^{144-154}$Sm.
Left panel: $E_x$ = 7-11 MeV, right panel: $E_x$ = 7-20 MeV. Red curves (ISS):
calculated with ISS-QRPA ($\Gamma$=0.1 MeV). Dotted blue curves (MM): QRPA
($\Gamma$=0.1 MeV) with the equilibrium deformations in
Tab.~\ref{equilibriumDef}. Green line (ISS+CD): ISS-QRPA with a 
energy-dependent width $\Gamma(E_x) = 2.5 (E_x/15)^2$ MeV. The black dots are
the measured $(\gamma,n)$ cross sections and the arrows labeled 
 by "n" mark the
position of the neutron-emission threshold of the respective isotope.}
\end{figure}

\section{Results}\label{results}

The results of ISS calculations for the  
 isotopic
 chains $^{78-86}$Kr,
$^{124-134}$Xe, $^{128-134}$Ba and $^{144-154}$Sm are shown in
Figs.~\ref{krypton}  -  \ref{samarium}. Our ISS  studies of $^{92-100}$Mo, 
$^{88}$Sr, and $^{90}$Zr are published in \cite{zha09}. The calculations for $^{139}$La are
published in \cite{Makinaga10}.
The figures show the calculations without collisional damping (denoted by ISS)
with a constant width $\Gamma$\,=\,0.1 MeV 
and with collisional damping (denoted by ISS+CD), where the latter were
obtained by folding the ISS cross section with the Breit-Wigner distribution
in Eq.~(\ref{eqn:cd}) with an energy-dependent width $\Gamma(E)$\,=\,2.5\,$(E/15)^2$ MeV.
For comparison we display also the QRPA results for the
equilibrium deformations, which are denoted by MM, where we use the values  in 
Table~\ref{equilibriumDef}, which were  calculated by M\"oller et al.
\cite{Moeller1,Moeller2} in the framework of their sophisticated Micro-Macro (MM) model.

 In order to characterize the strength function 
in a more global way we calculated  the   moments
\begin{equation}\label{mom}
m_k=\int_0^{E_u}dE E^k \left(S_{E1}(E)+S_{M1}(E)\right)
\end{equation}
 of our strength
functions $S_{E1,M1}(E)$, which are related by Eqs.~(\ref{sigmaE1},\ref{sigmaM1})
 to the cross sections shown in Figs. \ref{krypton} - \ref{samarium}. 
The centroid
energies $\bar E$, the widths $\bar\gamma$ and the integrated cross sections 
$\Sigma$  presented in Table~\ref{moments} are calculated by means of Eq. (\ref{mom}) from the
  ISS-QRPA results including the collisional
damping width $\Gamma(E)$, which are labelled as  ISS+CD in the figures. 
To characterize the  low-energy region the energy $E_u$ = 11 MeV is
considered as an appropriate upper integration limit. Concerning the entire GDR region the
moments are calculated using $E_u$ = 25 MeV as upper limit. The interval $E$ = 0-25 MeV 
 contains more than 85$\%$ of the Thomas-Kuhne-Reich (TRK) sum rule.
  Using MM-QRPA with collisional damping we also 
 obtained the same quantities for
the static equilibrium deformations, which are quoted in Table \ref{moments2}. 

Figs. \ref{krypton}-\ref{samarium}  show that the inclusion of CD damping (ISS+CD) eliminates
the fluctuations of 
 the ISS cross section. However it barely increases the 
average  cross section in the astrophysically interesting energy range, which was already 
found in our previous study of the Mo isotopes \cite{zha09}. Thus the dipole strength  
in this region results from the Landau fragmentation and the ISS fragmentation, which accounts  
for the various deformed shapes in the nuclear ground state.

In all considered chains there is a structural change from spherical 
to deformed shapes or reverse. The purely collective models for the GDR predict that an increasing
deformation leads to an increasing low-energy dipole cross section, because the GDR splits
into two separate peaks 
 \cite{Okomoto,Danos,sigmaBMII,DanosGreiner,jun08}. In Ref.~\cite{Doen07} the same relation between
deformation and low-energy dipole strength was found for the Mo 
 isotopes within the MM+QRPA
method based on a Nilsson potential, which was confirmed by
 ISS-QRPA calculations based on the WS potential \cite{zha09}.
This tendency  is also clearly seen for the Sm chain (cf.  Fig. \ref{samarium} 
and Tab. \ref{moments}).

\begin{table}
\caption{\label{moments} Integral properties derived from the moments $m_k$ of
the strength function $S_{E1}(E)$ in Eq.~(\ref{sigmaE1}) as calculated 
in ISS-QRPA with energy dependent width for the
even-$A$ series $^{78-86}$Kr,$^{130-136}$Ba,$^{124-136}$Xe and $^{144-154}$Sm: 
$\bar E=m_1/m_0$,  $\bar\gamma=\sqrt{(m_2/m_0)-(m_1/m_0)^2}$,  and the
integrated cross section $\Sigma = 16\pi^3/(3\hbar c)m_1$. The respective
integration limits $E =$ 11 MeV and $E =$ 25 MeV are indicated as an upper
index. The percentage of $\Sigma$ with respect to the TRK sum rule is given
in 
 the columns (\%).}
\begin{ruledtabular}
\begin{tabular}  
{rcccccccc}
$A$ & $\bar E^{11}$(MeV) & $\bar \gamma^{11}$(MeV) & $\Sigma^{11}$(MeV\,mb) &
$(\%)$ & $\bar E^{25}$(MeV)& $\bar\gamma^{25}$(MeV)& $\Sigma^{25}$(MeV\,mb) &
(\%) \\
\hline
Kr:  78 & 9.6 & 1.3 &  40\quad &  3.5 & 16.3 & 3.7 &  991 & 85 \\
     80 & 9.6 & 1.3 &  35 &  3.0 & 16.4 & 3.5 & 1024 & 86 \\
     82 & 9.6 & 1.2 &  35 &  2.9 & 16.4 & 3.4 & 1054 & 87 \\
     84 & 9.6 & 1.2 &  39 &  3.1 & 16.2 & 3.3 & 1095 & 89 \\
     86 & 9.5 & 1.2 &  45 &  3.6 & 15.9 & 3.3 & 1141 & 92 \\
        &     &     &     &      &      &     &      &    \\
Xe: 124 & 9.5 & 1.3 &  97 &  5.3 & 15.8 & 3.0 & 1599 & 87 \\
    126 & 9.5 & 1.3 &  98 &  5.3 & 15.1 & 3.3 & 1725 & 93 \\
    128 & 9.5 & 1.3 & 105 &  5.6 & 15.0 & 3.3 & 1750 & 93 \\
    130 & 9.5 & 1.3 & 109 &  5.7 & 15.0 & 3.3 & 1775 & 94 \\
    132 & 9.5 & 1.3 & 113 &  5.9 & 14.9 & 3.3 & 1806 & 94 \\
    134 & 9.5 & 1.3 & 120 &  6.2 & 14.8 & 3.2 & 1840 & 95 \\
        &     &     &     &      &      &     &      &    \\
Ba: 130 & 9.4 & 1.3 &  83 &  4.3 & 15.6 & 3.4 & 1746 & 91 \\
    132 & 9.4 & 1.3 &  82 &  4.2 & 15.6 & 3.3 & 1766 & 91 \\
    134 & 9.4 & 1.3 &  85 &  4.3 & 15.4 & 3.3 & 1803 & 92 \\
    136 & 9.4 & 1.3 &  88 &  4.5 & 15.4 & 3.2 & 1835 & 93 \\
    138 & 9.4 & 1.3 &  93 &  4.6 & 15.3 & 3.2 & 1866 & 93 \\
        &     &     &     &      &      &     &      &    \\
Sm: 144 & 9.4 & 1.3 & 95 &  4.5 & 15.8 & 2.9 & 1886 & 89 \\
    146 & 9.4 & 1.3 & 103 &  4.8 & 15.7 & 2.9 & 1907 & 89 \\
    148 & 9.4 & 1.3 & 110 &  5.1 & 15.7 & 2.9 & 1921 & 89 \\
    150 & 9.5 & 1.3 & 123 &  5.6 & 15.7 & 3.0 & 1923 & 88 \\
    152 & 9.5 & 1.4 & 182 & 8.3 & 15.7 & 3.1 & 1896 & 86 \\
    154 & 9.5 & 1.4 & 223 & 10.0 & 15.7 & 3.1 & 1892 & 85 \\
\end{tabular}
\end{ruledtabular}
\end{table}

\begin{table}
\caption{\label{moments2} Same integral properties as in Tab~\ref{moments} but here 
calculated in QRPA with the equilibrium deformations listed in Tab~\ref{equilibriumDef}.}
\begin{ruledtabular}
\begin{tabular} 
{rcccccccc}
$A$ & $\bar E^{11}$(MeV) & $\bar \gamma^{11}$(MeV) & $\Sigma^{11}$(MeV\,mb) &
$(\%)$ & $\bar E^{25}$(MeV)& $\bar\gamma^{25}$(MeV)& $\Sigma^{25}$(MeV\,mb) &
(\%) \\
\hline
Kr:  78 & 9.6 & 1.3 &  29 &  2.5 & 16.5 & 3.4 & 1000 & 86 \\
     80 & 9.7 & 1.1 &  33 &  2.8 & 16.6 & 3.2 & 1034 & 87 \\
     82 & 9.7 & 1.2 &  35 &  2.9 & 16.5 & 3.2 & 1059 & 87 \\
     84 & 9.6 & 1.1 &  40 &  3.3 & 16.2 & 3.2 & 1098 & 89 \\
     86 & 9.5 & 1.1 &  47 &  3.8 & 15.9 & 3.2 & 1144 & 91 \\
        &     &     &     &      &      &     &      &    \\
Xe: 124 & 9.5 & 1.3 &  96 &  5.3 & 15.2 & 3.3 & 1699 & 92 \\
    126 & 9.5 & 1.3 &  96 &  5.2 & 15.2 & 3.3 & 1728 & 93 \\
    128 & 9.5 & 1.3 & 103 &  5.6 & 15.1 & 3.3 & 1753 & 94 \\
    130 & 9.5 & 1.3 & 105 &  5.6 & 15.0 & 3.3 & 1778 & 94 \\
    132 & 9.5 & 1.2 & 105 &  5.5 & 15.0 & 3.2 & 1807 & 94 \\
    134 & 9.5 & 1.2 & 115 &  5.9 & 14.8 & 3.1 & 1841 & 95 \\
        &     &     &     &      &      &     &      &    \\
Ba: 130 & 9.4 & 1.3 &  76 &  4.0 & 15.6 & 3.3 & 1750 & 92 \\
    132 & 9.4 & 1.3 &  78 &  4.0 & 15.6 & 3.3 & 1771 & 92 \\
    134 & 9.4 & 1.3 &  81 &  4.2 & 15.5 & 3.2 & 1806 & 92 \\
    136 & 9.4 & 1.3 &  83 &  4.2 & 15.5 & 3.1 & 1835 & 93 \\
    138 & 9.4 & 1.3 &  90 &  4.5 & 15.3 & 3.1 & 1866 & 93 \\
        &     &     &     &      &      &     &      &    \\
Sm: 144 & 9.4 & 1.3 & 93 &  4.4 & 15.8 & 2.8 & 1890 & 89 \\
    146 & 9.4 & 1.3 & 98 &  4.6 & 15.8 & 2.8 & 1913 & 89 \\
    148 & 9.4 & 1.4 & 108 &  5.0 & 15.7 & 2.9 & 1923 & 89 \\
    150 & 9.4 & 1.4 & 118 &  5.4 & 15.7 & 3.0 & 1929 & 88 \\
    152 & 9.4 & 1.4 & 134 & 6.1 & 15.6 & 3.0 & 1942 & 88 \\
    154 & 9.4 & 1.5 & 151 & 6.8 & 15.5 & 3.0 & 1957 & 88 \\
\end{tabular}
\end{ruledtabular}
\end{table}

 According to the equilibrium deformations,
the lightest Kr isotope is oblate and all heavier ones have a nearly
spherical shape but are expected to be soft against deformation. 
The IBA parameters show a trend from $\gamma$-unstable ($\chi=0$) to spherical
($\zeta=0$), which is also visible in the deformation distributions in
Fig.~\ref{KrDistr}. 
 Although Fig.  \ref{krypton} indicates a certain narrowing
of the GDR with increasing $A$ (cf. right panel), this is not reflected by a decrease of the low-energy cross section 
in the left panel and the $\Sigma_{11}$ values in Tab. \ref{moments}, which are nearly constant.
In fact, the  $\Sigma_{11}$ values for the equilibrium deformation in Tab. \ref{moments2}  increase. 
There a two mechanisms to explain this unexpected behavior.   
One is the $A$-dependence of the GDR peak energy, which increases as  $A^{-1/3}$ along the isotopic chain.
The other can be traced to the emergence of resonances, which reflect
the bunching of particle-hole excitations due to the progressive degeneracy of the
single particle levels with decreasing deformation. The 
conspicuous    example is
the  strong peak near 10 MeV in the MM calculations,
seen in the left panel of Fig.  \ref{krypton}, which carries a summed strength 
of 22 mb MeV for $^{80}$Kr
and 26 mb MeV for $^{86}$Kr. The shape fluctuations in the ISS calculation
broaden it progressively with decreasing $A$, which reflects the increasing
probability of deformed shapes. A  flat bump remains of the resonance if CD is included.

According to Table~\ref{equilibriumDef}, the Xe
and Ba isotopes change from prolate through triaxially deformed to spherical 
shape with increasing neutron number, i.e. the deformation decreases with
the mass number $A$ not only in the Kr but also in Ba and Xe isotopes. From macroscopic
approaches that take into account the splitting of the GDR into two or three peaks caused by
axial or triaxial deformation, one expects  that  the values of
$\Sigma^{11}$ also decrease \cite{Okomoto,Danos,Thie83,jun08}. 
This trend is not visible. In all cases the values of 
$\Sigma^{11}$  increase in the MM calculations. 
According to the
IBA these isotopes  tend to  $\gamma$-instability as seen in
Fig.~\ref{triangles}. ISS also gives an increase of $\Sigma^{11}$ with $A$. 
Similar to the Kr isotopes, the MM results for Xe and Ba show strong
two-quasiparticle peaks for spherical shape, which are washed out by the shape fluctuations in
the ISS results. 
Only  the Sm chain, which spans the region  from spherical to prolate well deformed nuclei, 
shows the expected increase of  $\Sigma^{11}$ with $A$.
The examples demonstrate that the incompletely dissolved particle-hole structures can 
 substantially change the absorption cross section near the neutron threshold and can generate
 an $A$ dependence of the low-energy dipole strength that is opposite
 to the one of the macroscopic approaches. Nevertheless, all absorption 
 cross section, except the ones for semi-magic  nuclei, smoothly increase with energy.

One notices that for the considered isotopic chains there is only little experimental 
information. Concerning 
the $\gamma$-absorption cross sections at higher energies there exist $(\gamma,n)$ data 
for the Sm chain and for $^{138}$Ba \cite{Ripl2} but not 
for the other nuclides.   

The only $\gamma$-absorption cross sections for the low-energy 
region above 5 MeV up to the neutron emission threshold exist for $^{138}$Ba
from recent experiments with mono-energetic $\gamma$-rays at the HIGS
facility \cite{Ba138Tonchev}, which are included in Fig.~\ref{Barium}. 
The data for this  $N=82 $
semi-magic  nucleus show a bump at 8 MeV and possibly another at  9.5 MeV. 
The $N=82$ neighbor $^{139}$La was studied
in \cite{Makinaga10}. The data show a bump at 7 MeV and a shoulder around 9 MeV, 
which probably represent the same structures as in $^{138}$Ba.
The ISS+CD cross section  for  $^{138}$Ba has a shallow shoulder in this region.
In case of $^{139}$La \cite{Makinaga10}, the ISS+CD cross section has two very  broad peaks at  8.5 and 11.5 MeV. 
Comparing in Fig.~\ref{Barium} the ISS curve with the MM curve (zero deformation),  
one concludes that the structures originate from spherical two-quasiparticle excitations, 
which are strongly fragmented due to the shape fluctuations. The deviation
of their energies from experiment may indicate that our choice of the Wood-Saxon
potential does not quite correctly reproduce the single particle levels.   
A peak in the absorption cross section at 9 MeV  was also  was found  in the $N=50$ 
semi-magic   nuclides $^{88}$Sr, $^{89}$Y, and $^{90}$Zr \cite{88Sr-data} - \cite{89Y-data}. 
The ISS calculations \cite{zha09} give a peak at the correct energy, which is also a fragmented spherical
two-quasiparticle state. The inclusion of CD broadens the structure somewhat too strongly as compared
with the more pronounced "pygmy resonance" seen in experiment. Also in the case of the $N=82$
nuclides $^{138}$Ba and $^{139}$La, the combination ISS+CD seems to damp the spherical 
QRPA poles somewhat too strongly. Although the number of studied nuclei is still too small for
definite conclusions, one may speculate that the CD width depends stronger
on energy than assumed. The observation  that in many strongly deformed axial nuclei the  width of the
upper peak of the GDR is twice as big as the width of the lower peak might be taken as evidence for a strong increase
CD with energy. (However, as discussed below,  it may be caused by  Landau fragmentation as well).
The  quadratic energy dependence adopted in this paper is derived
from the schematic model of a Fermi gas. A function with a steeper energy dependence would give 
less CD in the threshold region if  its scale is adjusted to reproduce the peak height of the GDR.
Such a reduction of CD would barely  reduce the average absorption cross section at these energies
(compare the ISS and ISS+CD curves) but give more pronounced pygmy resonances.

In the right panels of Fig.~\ref{samarium}, the distance between the two GDR peaks is somewhat over estimated for the well deformed
isotopes $^{152,154}$Sm. The discrepancy can be traced to the large values of the 
deformation parameter $\beta$ of 0.306 and 0.341, respectively, which reflect 
the large 
experimental $B(E2,2^+\rightarrow0^+)$ values \cite{raman}. The estimate of 
the hydrodynamic model
 $2\left (E(K=1)-E(K=0))/(E(K=1)+E(K=0)\right)=0.94\beta$ 
 \mbox{ \cite{Okomoto,Danos,sigmaBMII,DanosGreiner,jun08}}
gives similar splittings of 4.1 and 4.5 MeV, respectively. This indicates some  inconsistency 
between the experimental $B(E2,2^+\rightarrow0^+)$ value and the observed splitting
of the GDR. Another problem is visible for the well deformed
isotope $^{154}$Sm. In experiment, the second GDR peak has a somewhat smaller height
than the first one, whereas  in the calculation it is opposite.
Since the high-energy peak is two-fold degenerated ($K=\pm 1$) it 
carries twice the strength  of 
  the non-degenerate ($K=0$) low-energy peak.
 In order to be lower, its width must be 
more than two times the width of the first peak. However, collisional damping 
$\Gamma \propto E_x^2$ gives only $~(16/12)^2=1.8$.  The stronger Landau 
fragmentation of the low-energy peak acerbates  the discrepancy
in the considered case of  $^{154}$Sm. Many of the well deformed 
nuclei behave in the same way: The high-energy peak has the same
height as the low-energy one, indicating that its width must be about twice
\cite{Dietrich88}. Different versions of the mean field produce 
different pattern of the Landau fragmentation. In the calculations of Ref. \cite{skyrme-separ2}
for $^{154}$Sm, the 
Skyrme density functional SLy6 produces stronger Landau fragmentation
for the $K=\pm1$ peak than for the $K=0$ peak, resulting in a better agreement 
with experiment, whereas the functionals SkM*, Skl3, and SkT6 give a too high second maximum.  
For the same nucleus, Ref. \cite{skyrme-full} obtains 
sufficient Landau fragmentation in the upper peak for the Skyrme functional SKM*,
however not for SLy4 and SkP.  The results for $^{154}$Sm obtained 
in \cite{skyrme-separ2} and \cite{skyrme-full} with SkM* disagree. At this point it seems
unclear if Landau fragmentation  can account for  
the widths of the two GDR peaks observed in well deformed axial nuclei.

\section{Range of validity  of ISS}\label{range}

The coupling between the low-frequency quadrupole mode and the purely
collective GDR mode was studied in
Ref.~\cite{DanosGreiner} by means of the Dynamic Nuclear Collective Model (DNCM). 
The authors find that the dipole strength becomes
distributed over several quadrupole excitations. The  collisional damping
washes out the discrete structures to a smooth envelop.  In
Ref.~\cite{LeTourneux} the validity of  the ISS approximation was investigated
in the same model. The discrete spectrum of the DNCM was compared
with the continuous strength function obtained by integrating the 
instantaneous excitation probabilities of the GDR over the probability
distribution of the shape parameters in the ground state, which corresponds to
a dense set of sampling points in ISS.
 The resulting smooth strength function  becomes a good approximation of the 
envelope
 of the discrete lines if
\begin{equation}\label{xi}
\xi=\frac{d\omega_1}{d\beta}\frac{ \beta_0}{\omega_2}\gg 1,
\end{equation}
 where $\beta_0$
is the zero point amplitude of the quadrupole vibration and $\omega_1$ and $\omega_2$ are
the frequencies of the dipole and quadrupole vibrations, respectively. 
In our preceding paper \cite{zha09},  we gave some qualitative estimate that
the ISS is applicable to the dipole excitations around the particle emission thresholds according to this condition.

In order to judge the quality of the ISS approximation in this energy region in more detail, we
 studied the coupling of  
a single QRPA $1^-$ 
pole with the collective quadrupole 
$(2^+)$
mode in a schematic model. 
Since the $1^-$ 
poles of interest are located substantially away
from the peak of the GDR, it suffices for the following discussion to 
consider them as dressed two-quasiparticle excitations. Accordingly, their
energy  is not very different from the  two-quasiparticle energy,
and their 
transition strength is given by the two-quasiparticle strength times an effective charge accounting for the
screening. (A detailed discussion of this approximation is given in \cite{eeffectiveBMII}.)  
Thus, it is sufficient to use the energies and transition strengths of the 
two-quasiparticle excitations 
in the threshold region as
a starting point for the following schematic model.  

 We generated all two-quasiparticle 
$1^-$ 
excitations in the energy interval between 7 and 8 MeV
 and studied their dependence on deformation. We followed 
the development in a diabatic way, such that we calculated the overlap of the 
wave functions of the quasiparticle states at 
adjacent deformations 
$\beta_n$ and $\beta_{n+1}$ 
and associated $1^-$ states 
by requiring maximal overlap between $n$ and $n+1$ (cf. \cite{TAC}).
A typical change of the 
two-quasiparticle 
energy was found  to be 1-3 MeV 
over the interval $0.1\leq \beta \leq 0.3$.  
One such two-quasiparticle state is selected and its coupling 
to an axial collective 
quadrupole degree of freedom
 is considered,  assuming that the coupling
is caused by the deformation dependence two-quasiparticle energy (as in the DNCM).  
The quadrupole mode is described by 10 equidistant points
$\beta_n=0.04\,n, ~n=1, .., 10$  and $\gamma=0$. The step size is chosen to
roughly agree with the distance between the samples in the realistic ISS calculations (cf. Fig. ~\ref{KrDistr}).  
 The Hamiltonian for the collective 
quadrupole 
motion  is given by the matrix 
 \begin{equation}
 H^{(c)}_{n,m}=h\left(\delta_{n,m+1}-2\delta_{n,m}+\delta_{n,m-1}\right)+V(m)\delta_{n,m}~.
 \end{equation}
 The first term is the discretized  kinetic energy 
where $h$ determines the inertial mass parameter. 
The second is the discretized potential.
We studied the two potentials 
 \begin{eqnarray}
 V_{\rm HO}(n)&=& D(n-7)^2, \\
V_{\rm SW}(n) &=& \left\{ \begin{array}{r@{\quad {\rm for} \quad}l}
                                0 & n = 3,~...,~8  \\
                                \infty &  n=1,~2,~9,~10 
                                \end{array} \right.,
\end{eqnarray}
which are the discrete versions of a harmonic oscillator and a square-well potential, respectively.
The two-quasiparticle energy is taken as $E^{(2qp)}(n)=(7+ \Delta e~(n-7))$ MeV
 with $\Delta e~=0.2$ and 0.6, corresponding to a change of 1 and 3 MeV over the considered deformation range, 
 respectively.
The Hamiltonian describing the two-quasiparticle state coupled to the quadrupole mode is
\begin{equation}
 H^{(c,2qp)}_{n,m}= H^{(c)}_{n,m}+(7+\Delta e(n-7))\delta_{n,m}.
 \end{equation}
  Both $H^{(c)}$ and  $ H^{(c,2qp)}$ are diagonalized numerically.
 The resulting eigenvalues are $ E^{(c)}(i)$ and $E^{(c,2qp)}(j)$.
  The resulting eigenvectors are $  U^{(c)}_{n,i}$ and $ U^{(c,2qp)}_{n,j}$, respectively, 
   where $i$ and $j$ label the respective eigenstates. 
  The transition strength from the ground state $1$ to the mixed excited state $j$ at the 
  energy $E^{(c,2qp)}(j)$ is 
  \begin{equation}\label{sex}
  S_{ex}(j)=\left[\sum_n s(n) U^{(c)}_{n,1} U^{(c,2qp)}_{n,j}\right]^2,
   \end{equation}
 where $s(n)$ is the 2qp transition matrix element for deformation point $\beta_n$.
 The analogous ISS strength for the sampling point $n$ is
 \begin{equation}\label{siss}
  S_{ISS}(n)=\left[ s(n) U^{(c)}_{n,1}U^{(c)}_{n,1}\right]^2,
   \end{equation}   
   which is associated with the energy $E^{(2qp)}(n)$. Strength functions 
   are generated by folding with a Breit-Wigner distribution in order
   to account for CD,
   \begin{eqnarray}
  S_{ex}(E)=\sum_j S_{ex}(j) \frac{\Gamma}{2\pi\left((E^{(c,2qp)}(j)-E)^2+(\Gamma/2)^2\right)},\label{sfex}\\
  S_{ISS}(E)=\sum_n S_{ISS}(n) \frac{\Gamma}{2\pi\left((E^{(2qp)}(n)-E)^2+(\Gamma/2)^2\right)}\label{sfiss}.  
   \end{eqnarray}   

\begin{figure}\label{hop}
\includegraphics[width=8cm]{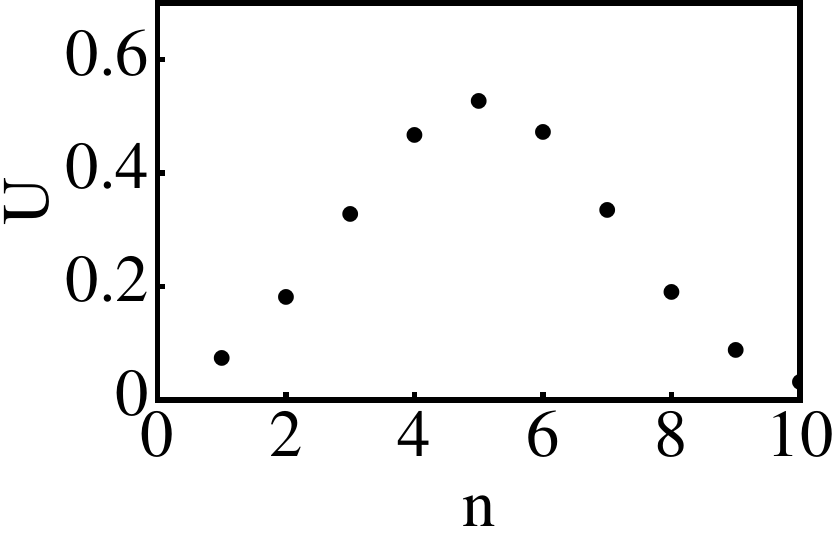}
\includegraphics[width=8cm]{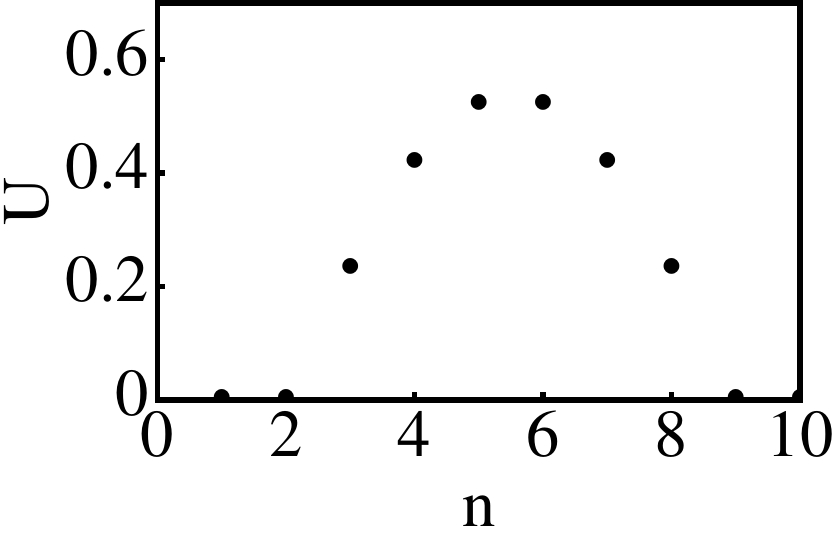}
\caption{\label{wf}Ground state eigenvectors 
for the harmonic oscillator (left) and square well (right). }
\end{figure}
\begin{figure}\label{hopstrength}
\includegraphics[width=8cm]{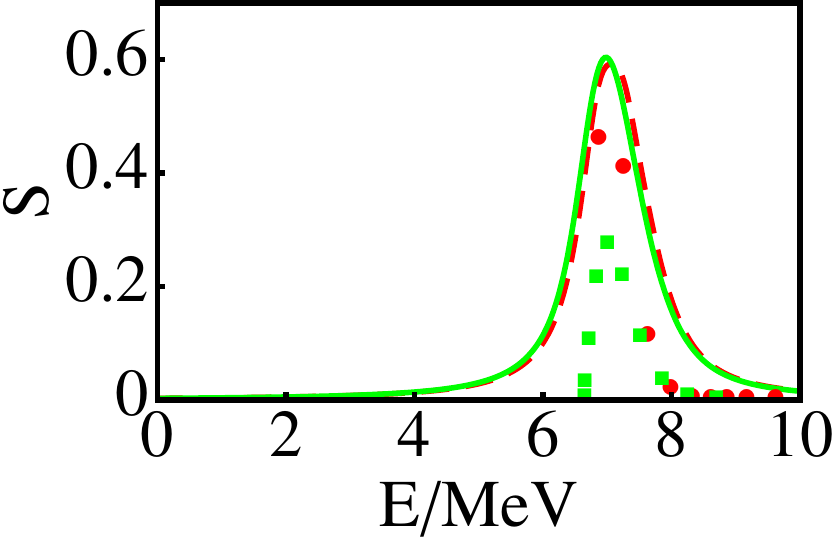}
\includegraphics[width=8cm]{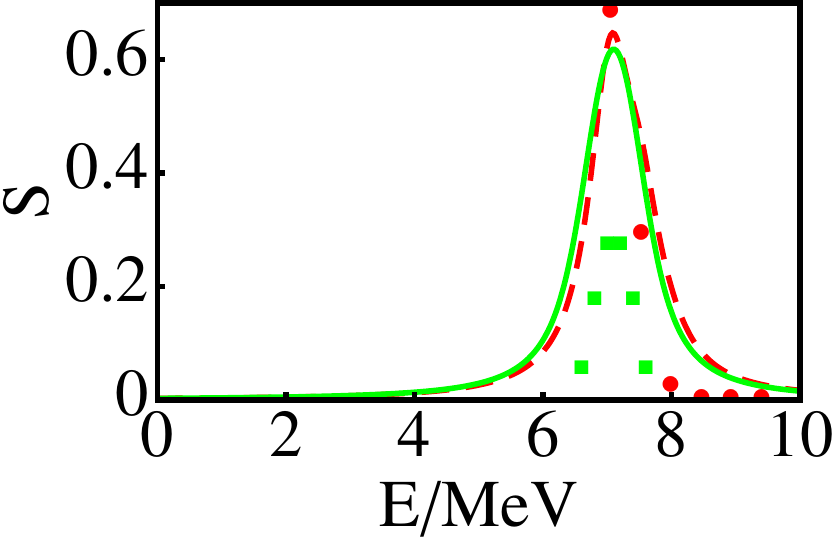}
\includegraphics[width=8cm]{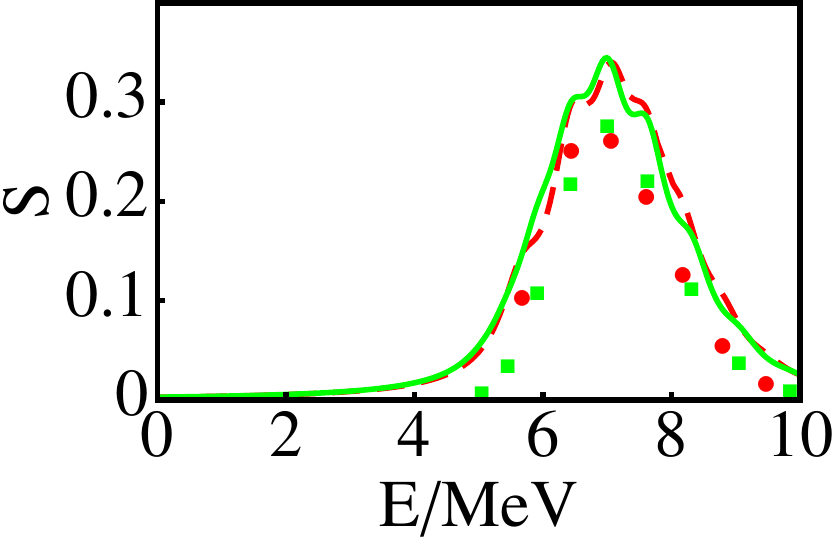}
\includegraphics[width=8cm]{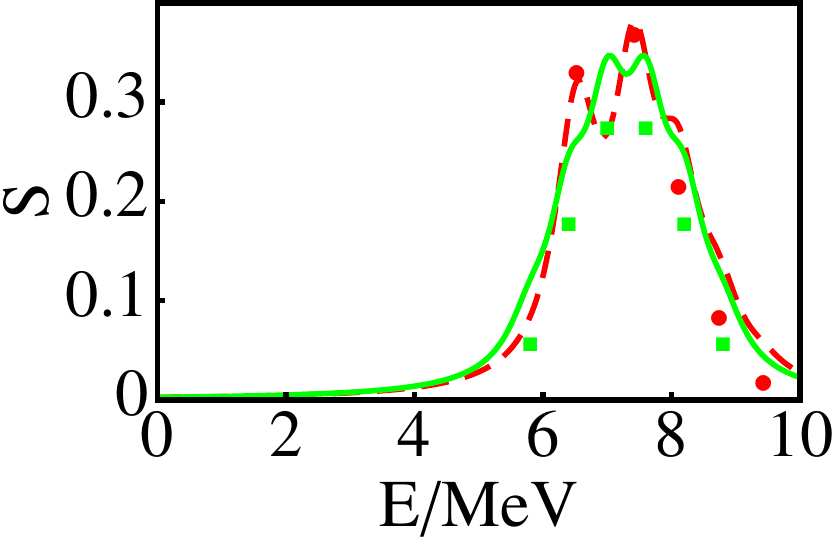}
\caption{\label{hopping}(Color online) Strength functions  of a RPA pole
coupled to the collective quadrupole mode. Red circles - exact, 
red curve dashed - exact+CD ($\Gamma$ = 0.8 MeV), green squares - ISS, 
green curve - ISS+CD ( $\Gamma$ = 0.8 MeV). Change of the pole  energy 
between $\beta$=0.1 and 0.3:  upper panels - 1 MeV,  lower panels - 3 MeV.
Potentials: left panels - harmonic oscillator, right panels - square well. }
\end{figure}

In the following we discuss only the results for  $s(n)=1$, because the  study of  non-constant $s(n)$ values, 
which were derived from 
 the two-quasiparticle excitations in the same way as the two-quasiparticle energies,   led to the same conclusions. 
The choice $h=1.12$MeV and $D=0.028$MeV gives an excitation energy of $E^{(c)}(2)-E^{(c)}(1)$ = 0.31 MeV 
for the collective quadrupole mode.
 The corresponding
ground-state eigenvectors, shown in Fig.~\ref{wf}, are distributed over 5 sampling points.  Fig.~\ref{hopping} compares 
the exact transition strength (\ref{sex}) with the ISS approximation (\ref{siss}). In the case of the harmonic oscillator
potential, the oscillator length is $3\Delta \beta=0.12$.  In the upper panel the two-quasiparticle energy changes by
1 MeV between $\beta=0.1$ and 0.3 ($n=3$ and 8, respectively), which corresponds to  $\xi=2$ (cf. Eq. (\ref{xi})). 
As seen, the energies of the coupled states do not agree with the energies of the sampling points.   Nevertheless,
the locations and the widths of the distributions are roughly the same. In the lower panels the two-quasiparticle energy
changes by 3 MeV between $\beta=0.1$ and 0.3, which corresponds to  $\xi=6$. Now ISS becomes a rather
good approximation to the exact transition strength. This agrees with  Ref. \cite{ LeTourneux}. Using  the
continuous version of ISS, the authors found that for $\xi>5$ the strong coupling limit is approached and substantiates
the  discussion in our previous paper \cite{zha09}.  In the case of the square-well potential, the strong coupling limit
is somewhat slower approached. Presumably this reflects the fact that
  eigenvectors have a lower probability at the turning than for the harmonic oscillator. At the turning point,
 the energy equals the potential energy, which is assumed in ISS. 
 
 The strength functions take CD into account by folding the transition strengths with a Breit-Wigner distribution. The width
 $\Gamma=0.8$ MeV corresponds to the energy-depend width at 7 MeV, used throughout this paper.   In the 
 upper panels of Fig. \ref{hopstrength}, the 
 energy difference between the sample points  is much smaller than the damping width, $\Delta E/\Gamma=0.2$.
  All structure is averaged out. ISS and the exact strength function are practically identical.
 In the lower panels of Fig. \ref{hopstrength} the results for $\Delta E/\Gamma=0.6$ are displayed. 
 Some of the structure survives. In the case of the harmonic 
 oscillator the ISS and exact strength functions nearly agree, because the discrete transition strengths are very similar.
 In the case of the square-well potential the surviving structures disagree.  
 Since the coupling of the two-quasiparticle state to the collective mode is weaker, the energies and
 transitions strengths of the mixed states disagree with those of the sampling points, which is transfered
 to the strength functions. Hence the fluctuations of the ISS strength function
  do not represent physical structures. They are just "sampling noise" that should be disregarded. 
  The ISS strength functions
  for most of the nuclides studied in this paper do barely show structure around the particle emission thresholds,
  which means that ISS is reliable. The shallow peaks of the ISS+CD strength functions  in the Kr isotopes and
  Sr and Zr isotopes \cite{zha09} should be considered  as real structures ("pygmy resonances") predicted
  by our model, because their width is much larger than the CD damping width.
  
  The number of sampling points is determined by the number of bosons used for  diagonalizing the IBA Hamiltonian. In order
  to study the effects of this coarse graining of the collective mode, we decreased the deformation
  step in our  schematic model to $\Delta \beta=0.02$. The results for the exact strength functions changed
  only marginally, which means it is sufficient to restrict the number of bosons to 10. The ISS sampling noise
  is suppressed on this finer grid. All ISS strength function become smooth peaks because $\Delta E/\Gamma=0.3$. 
    The fine structure of the ISS  strength function in
  the left lower panel of Fig.~\ref{hopping} is similar to the  exact one, because accidentally we chose
  the sample points  near the  deformation points where  the mixed states localize.
  In the case of a spherical nucleus and/or at sufficiently low excitation energy, where CD is weak, one expects observing 
  the mixed states as resolved lines. Obviously, ISS will not describe these lines individually. Still, the location and the width
  of the distribution of lines (representing sampling points) will correlate with the location and width
  of a fragmented QRPA pole.

\section{Conclusions}\label{conc}

The method of Instantaneous Shape Sampling (ISS), suggested in our preceding 
Rapid Communication \cite{zha09} has been presented in extended form.
It relies on the assumption that the photo excitation is a fast process as compared to 
the shape fluctuations of nuclei, such that the total $\gamma$-absorption cross section
is the sum of the absorption cross sections of a set of instantaneous shapes, each weighted 
with its  probability being present in the ground state. That is, the  $\gamma$-quant 
"takes a snapshot of the instantaneous shape of the nucleus" when being absorbed.  
In the present implementation of the ISS concept, the quadrupole motion is described by
the Interacting Boson Approximation and the $\gamma$-absorption is described by
the Quasiparticle Random Phase Approximation(QRPA)
 for a deformed Woods Saxon potential
combined with a dipole  -  dipole interaction for the $E1$ modes and a spin-spin interaction for the 
$M1$ modes. The ISS concept can be applied to other versions of QRPA and a different
description of the collective quadrupole mode.    

Studying the coupling between a dipole QRPA solution near the particle threshold 
and the low-lying quadrupole mode
in a schematic model, we found that ISS provides a good description of the location and the
width of the resulting group of levels, although there is no one-to-one correspondence between
the lines. Taking into account the collisional damping (CD) by folding the discrete QRPA solutions
 with a Breit-Wigner function, 
the resulting ISS+CD strength functions reproduce the exact ones very well,
if the distance between the energies of the coupled states is smaller than the damping width.  

We applied our version of ISS-QRPA to the chains of the Kr, Ba, Xe, and Sm isotopes, which all 
span the transitional regions between deformed and spherical shape. As in our previous study of the
Mo isotopes \cite{zha09}, we find that the dipole absorption cross section in 
the energy region of photo-nuclear reactions is determined by the Landau fragmentation
and the dynamical deformation. In order to reproduce
 the broad  peak of the Giant Dipole Resonance  (GDR) additional CD must be 
introduced, which we assumed to be proportional to the square of the photon energy. 
Its scale turned out to be nearly independent of 
 the nuclear mass.
CD smoothes out most of the fluctuations of the ISS-QRPA 
absorption cross section, whereby it does not increase the cross section in
the  energy region of photo-nuclear reactions in any substantial way. 
For all but semi-magic nuclei, the resulting absorption cross  section increases with energy in a smooth way,
as observed. 

Collective hydrodynamic descriptions of the GDR 
give an increase of the low-energy dipole absorption cross section with nuclear
deformation. It is caused by the splitting of the GDR into a low frequency oscillation along the long and two 
high frequency oscillations along the short axes.
In the case of the Mo and Sm isotopic
chains, the deformation increases  with neutron number. ISS-QRPA reproduces the expected 
increase of the low-energy
dipole absorption cross section.  However, in case of  the Kr, Ba, and Xe chains,
for which the deformation decreases  with neutron number, 
the expected decrease of the low-energy cross section 
is not found. ISS-QRPA predicts a nearly constant value
of the cumulative low-energy cross section ($E_x\leq 11$ MeV)  
when approaching the shell closure. The reason is the $A^{-1/3}$ decrease of the GDR peak energy 
as well as the progressive bunching of the two-qpasiparticle 
excitations when approaching spherical shape.

In the case of semi-magic nuclei, relicts of these  bunches
 survive the damping
by shape fluctuations and collisional damping. They appear as broad bumps in 
ISS-QRPA cross section,
which may substantially enhance the absorption cross section around the particle
thresholds. These "pygmy resonances" 
are two-quasiparticle excitations dressed with isovector dipole vibrations and fragmented by
coupling to shape fluctuations. In the case of the $N=50$ isotones, the position
of the resonance is well reproduced. In the case of the $N=82$ isotones,
some discrepancy between the calculated and observed location may point to
inaccuracies of the single particle levels  of the adopted Woods-Saxon potential.

Acknowledgements: This work was supported by the German DFG project KA2519/1-1 and the
US DOE grant DE-FG02-95ER4093.

\end{document}